%% file: KT19_asymmetric.tex
\documentclass[11pt,letterpaper,twoside,reqno,nosumlimits]{amsart}

\input{KT19_macro}

\allowdisplaybreaks

\title[Asymmetric binary component decomposition]{Binary component decomposition \\ Part II: The asymmetric case}

\author[R.~Kueng]{Richard~Kueng}
\author[J.~A.~Tropp]{Joel~A.~Tropp}

\date{31 July 2019}



\begin{document}

\begin{abstract}
This paper studies the problem of decomposing a low-rank
matrix into a factor with binary entries,
either from $\{\pm 1\}$ or from $\{0,1\}$,
and an unconstrained factor.
The research answers fundamental questions about the existence
and uniqueness of these decompositions.  It also leads to tractable
factorization algorithms that succeed under a mild deterministic
condition.
This work builds on a companion paper that addresses
the related problem of decomposing a low-rank positive-semidefinite matrix
into symmetric binary factors.
\end{abstract}

\maketitle

\section{Motivation}

Constrained matrix decompositions are among the basic methods for
unsupervised data analysis.  These techniques play a role in
many scientific and engineering fields,
ranging from environmental engineering~\cite{PT94:Positive-Matrix} and neuroscience~\cite{OF96:Emergence-Simple-Cell}
to signal processing~\cite{Com94:Independent-Component} and statistics~\cite{ZHT06:Sparse-PCA}.
Constrained factorizations are powerful tools
for identifying latent structure in a matrix;
they also support data compression, summarization, and visualization.

The literature contains a number of frameworks~\cite{TB99:Probabilistic-Principal,CDS02:Generalization-Principal,Tro04:Phd-Thesis,Sre04:Learning-Matrix,Wit10:Phd-Thesis,Jag11:Phd-Thesis,Bac13:Convex-Relaxations,Ude15:Phd-Thesis,BE16:Hierarchical-Compound-Poisson-Factorization,Bru17:Phd-Thesis,HV19:Low-Rank}
for thinking about constrained matrix factorization
and for developing algorithms that pursue these factorizations.
Nevertheless, we still lack theory that fully justifies these approaches.
For instance, researchers have only attained a partial understanding
of which factorization models are identifiable and which ones we can
compute provably using efficient algorithms.

The purpose of this paper and its companion~\cite{KT19:Binary-Factorization-I}
is to develop foundational results on 
factorization models that we call \emph{binary component decompositions}.
In these models, one (or both) of the factors takes values in the set $\{ \pm 1 \}$ or in the set $\{0, 1\}$.
Binary component decompositions are appropriate when the latent factors reflect an exclusive choice.
From a mathematical perspective, these constrained factorizations also happen
to be among the easiest ones to understand.

In this second paper, we consider the problem of factorizing a rectangular matrix
into a binary factor and an unconstrained matrix of weights.  We develop
results on existence, uniqueness, tractable computation, and robustness
to gross errors.  Our analysis builds heavily on the work in the
companion paper~\cite{KT19:Binary-Factorization-I}, which treats the
problem of decomposing a positive-semidefinite matrix into symmetric
binary factors.

\subsection{Notation}

We rely on standard notation from linear algebra and optimization.
Scalars are written with lowercase Roman or Greek letters ($x, \xi$);
lowercase bold letters ($\vct{x}, \vct{\xi}$) denote (column) vectors;
uppercase bold letters ($\mtx{X}, \mtx{\Xi}$) denote matrices.
We reserve calligraphic letters ($\mathcal{X}$) for sets.
The symbol $\lesssim$ suppresses universal constants.

Throughout, $n$ and $m$ are natural numbers.
We work in the real linear spaces $\R^n$ and $\R^m$ equipped with the
standard inner product $\ip{\cdot}{\cdot}$ and the associated norm topology featuring $\| \vct{x} \|_{\ell_2} = \sqrt{\langle \vct{x},\vct{x}\rangle}$.
The standard basis vector $\mathbf{e}_i$ has a one in the $i$th coordinate
and zeros elsewhere, while $\mathbf{e}$ is the vector of ones;
the dimension of these vectors depends on context.
The map ${}^\transp$ transposes a vector or matrix.
The binary operator $\odot$ is the Schur (i.e., componentwise) product of vectors.
The closed and open probability simplices are the sets
\begin{displaymath}
\Delta_r = \left\{ \vct{\tau} \in \R^r : \text{$\tau_i \geq 0$ and $\sum_{i=1}^r \tau_i = 1$} \right\}
\quad\text{and}\quad
\Delta_r^+ = \left\{ \vct{\tau} \in \R^r : \text{$\tau_i > 0$ and $\sum_{i=1}^r \tau_i = 1$} \right\}.
\end{displaymath}

We write $\Sym_n$ for the linear space of symmetric $n \times n$ real matrices.
The symbol $\Id$ denotes the identity matrix, and $\mathbf{E}$ denotes the matrix of ones;
their dimensions are determined by the context.
The dagger ${}^\dagger$ refers to the Moore--Penrose pseudoinverse.
A positive-semidefinite (psd) matrix is a symmetric matrix $\mtx{X}$ that satisfies
$\vct{u}^\transp \mtx{X} \vct{u} \geq 0$ for all vectors $\vct{u}$ with compatible dimension.
The statement $\mtx{X} \psdge \mtx{Y}$ means that $\mtx{X} - \mtx{Y}$ is psd,
and $\mtx{X} \psdgt \mtx{Y}$ means that $\mtx{X} - \mtx{Y}$ is strictly positive definite, i.e. $\vct{u}^\transp (\mtx{X}-\mtx{Y}) \vct{u} >0$ for all vectors $\vct{u}$ with compatible dimension.

\section{Sign component decomposition and binary component decomposition}

We begin with a short discussion of the singular-value decomposition
and its properties (Section~\ref{sec:svd}).  Afterward, we introduce the two factorizations that
we treat in this paper, the sign component decomposition (Section~\ref{sec:scd})
and the binary component decomposition (Section~\ref{sec:bcd-intro}).
We present our main results on situations where these factorizations
are uniquely determined and when they can be computed using efficient
algorithms.  An outline of the rest of the paper appears in Section~\ref{sec:roadmap}.

\subsection{The singular-value decomposition}
\label{sec:svd}

We begin with the singular-value decomposition (SVD),
the royal emperor among all matrix factorizations.
Let $\mtx{B} \in \R^{n \times m}$ be a rectangular matrix.
For some natural number $r \leq \min\{m, n\}$,
we can decompose this matrix as
\begin{equation} \label{eqn:svd-vector}
\mtx{B} = \sum_{i=1}^r \sigma_i \vct{u}_i \vct{v}_i^\transp.
\end{equation}
In this expression, $\{ \vct{u}_1, \dots, \vct{u}_r \} \subset \R^n$
and $\{ \vct{v}_1, \dots, \vct{v}_r \} \subset \R^m$ are orthonormal
families of left and right singular vectors associated with the positive singular values
$\sigma_1 \geq \sigma_2 \geq \dots \geq \sigma_r > 0$.
We can also convert the decomposition~\eqref{eqn:svd-vector}
into a matrix factorization:
\begin{equation} \label{eqn:svd-matrix}
\mtx{B} = \mtx{U} \, \mtx{\Sigma} \, \mtx{V}^\transp
\quad\text{where}\quad
\begin{aligned}
\mtx{U} &= \begin{bmatrix} \vct{u}_1 & \dots & \vct{u}_r \end{bmatrix} \in \R^{n \times r}; \\
\mtx{\Sigma} &= \diag(\sigma_1, \dots, \sigma_r) \in \R^{r \times r}; \\
\mtx{V} &= \begin{bmatrix} \vct{v}_1 & \dots & \vct{v}_r \end{bmatrix} \in \R^{m \times r}. \\
\end{aligned}
\end{equation}
The matrices $\mtx{U}$ and $\mtx{V}$ are orthonormal; that is, $\mtx{U}^\transp \mtx{U} = \Id$
and $\mtx{V}^\transp \mtx{V} = \Id$.

The singular-value decomposition is intimately connected to the problem
of finding a best low-rank approximation of a matrix~\cite{Mir60:Symmetric-Gauge}.
Indeed, for any unitarily invariant norm $\norm{\cdot}$,
\begin{displaymath}
\min_{\rank \mtx{L} = k} \norm{ \mtx{B} - \mtx{L} }
	= \norm{ \mtx{B} - \sum_{i=1}^k \sigma_i \vct{u}_i \vct{v}_i^\transp }
	\quad\text{for each $k = 1, \dots, r$.}
\end{displaymath}
This variational property has a wide range of consequences,
both theoretical and applied.

The singular-value decomposition also holds a distinguished place in statistics
because of its connection with principal component analysis~\cite{Jol02:Principal-Component}.
Given a data matrix $\mtx{B} \in \R^{n \times m}$ with standardized%
\footnote{A vector is \emph{standardized} if its entries sum to zero
and its Euclidean norm equals one.}
rows, we can perform a singular-value decomposition
to express $\mtx{B} = \mtx{UW}^\transp$, where $\mtx{W} = \mtx{\Sigma} \mtx{V}^\transp$.
In this setting, the left singular vectors $\vct{u}_i$ are called
\emph{principal components}, the directions in
which the columns of $\mtx{B}$ exhibit the most variability.
The entries of the matrix $\mtx{W}$ are called \emph{weights} or \emph{loadings};
they are the coefficients with which we combine the principal components to
express the original data points.

On the positive side of the ledger,
the singular-value decomposition~\eqref{eqn:svd-vector}--\eqref{eqn:svd-matrix} always exists,
and it is uniquely determined when the (nonzero) singular values are distinct.
Moreover, we can compute the singular-value decomposition, up to a fixed (high) accuracy,
by means of highly refined algorithms, in polynomial time.

On the negative side, we cannot impose constraints on the singular vectors
to enforce prior knowledge about the data.  Second, we generally cannot assign
an interpretation or meaning to the singular vectors, without committing the
sin of reification.  Moreover, the orthogonality of singular vectors may
not be an appropriate constraint in applications.  Structured matrix factorizations
are designed to address one or more of these shortcomings.

\subsection{Sign component decomposition}
\label{sec:scd}

In this project, we consider matrix factorization models where one of the
factors is required to take binary values.
In this section, we treat the
case where the entries of the binary factor are limited to the set $\{ \pm 1 \}$.
In Section~\ref{sec:bcd-intro}, we turn to the case where the entries
are drawn from the set $\{ 0, 1 \}$.

\subsubsection{The decomposition}

\begin{figure}[t]
\begin{tikzpicture}[scale=0.6]
\draw[very thick,fill=blue!10] (0,0) rectangle (5,4);
\node at (4.5,3.5) {\Large $\mtx{B}$};
\node at (6,2) {\Large $=$};
\draw[very thick,fill=blue!10] (7,0) rectangle (9,4);
\node at (8.5,3.5) {\Large $\mtx{S}\phantom{{}^\transp\!\!}$};
\node at (8,2) {\Large $\pm$};
\draw[very thick,fill=blue!10] (9.5,2) rectangle (14.5,4);
\node[right] at (9.5,3.5) {\Large $\mtx{W}^\transp$};
\node at (15.5,2) {\Large $=$};
\node[right] at (16,2) {\Huge $\sum_i$}; 
\draw[very thick,fill=blue!10] (18.5,0) rectangle (19.5,4);
\node at (19,3.5) {\Large $\vct{s}_i\phantom{{\vct{s}_i}^\transp\hspace{-6mm}}$};
\node at (19,2) {\Large $\pm$};
\draw[very thick,fill=blue!10] (20,3) rectangle (25,4);
\node[right] at (20,3.5) {\Large ${\vct{w}_i}^\transp$};
\end{tikzpicture}
\caption{\textit{Asymmetric sign component decomposition.}
The sign component decomposition~\eqref{eqn:scd-matrix}--\eqref{eqn:scd-vector}
expresses a rectangular matrix $\mtx{B}$ as the product of a sign matrix $\mtx{S}$
and an unconstrained matrix $\mtx{W}^\transp$.} \label{fig:scd}
\end{figure}
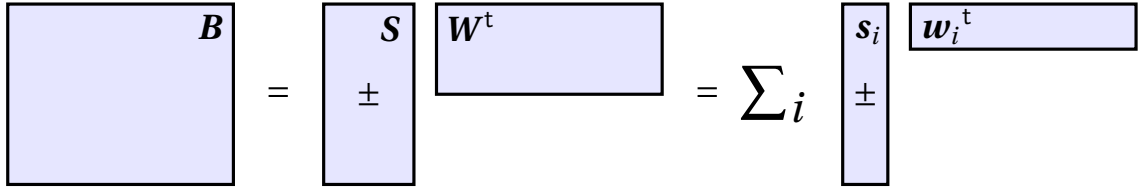

As before, assume that $\mtx{B} \in \R^{n \times m}$ is a rectangular matrix.
We seek a decomposition of the form
\begin{equation} \label{eqn:scd-matrix}
\mtx{B} = \mtx{SW}^\transp
\quad\text{where}\quad
\mtx{S} \in \{ \pm 1 \}^{n \times r}
\quad\text{and}\quad
\mtx{W} \in \R^{m \times r}.
\end{equation}
This factorization can also be written in vector notation as
\begin{equation} \label{eqn:scd-vector}
\mtx{B} = \sum_{i=1}^r \vct{s}_i \vct{w}_i^\transp
\quad\text{where}\quad
\vct{s}_i \in \{ \pm 1 \}^{n}
\quad\text{and}\quad
\vct{w}_i \in \R^m.
\end{equation}
We call~\eqref{eqn:scd-matrix}--\eqref{eqn:scd-vector} an (asymmetric) \emph{sign component decomposition}
of the matrix $\mtx{B}$.  The left factor $\mtx{S}$ is called the \emph{sign component};
its columns $\vct{s}_i$ are also called \emph{sign components}.  The right factor
$\mtx{W}$ is unconstrained; its entries are called \emph{weights} or \emph{loadings}.
See Figure~\ref{fig:scd} for an illustration.

It is not hard to show that each $n \times m$ matrix $\mtx{B}$ admits a plethora of distinct
sign component decompositions~\eqref{eqn:scd-matrix} where the inner dimension is $r = n$;
see Proposition~\ref{prop:existence}.
It is more interesting to consider a low-rank matrix $\mtx{B}$
and to search for \emph{minimal} decompositions,
those where the inner dimension $r$ of the factorization~\eqref{eqn:scd-matrix}
equals the rank of $\mtx{B}$.

\begin{remark}[Matrix sign function]
The sign component decomposition must not be confused with
the matrix sign function, which is a spectral computation
related to the polar factorization~\cite[Chap.~5]{Hig08:Functions-Matrices}.
\end{remark}

\subsubsection{Schur independence}

The sign component decomposition~\eqref{eqn:scd-matrix}--\eqref{eqn:scd-vector}
has a combinatorial quality, which suggests that it might be hard to find.
Remarkably, there is a large class of matrices for which we can tractably
compute a minimal sign component decomposition.
The core requirement is that the
sign components must be somewhat different.
The following definition~\cite{LP96:Facial-Structure,KT19:Binary-Factorization-I}
encapsulates this idea.

\begin{definition}[Schur independence of sign vectors] \label{def:schur-independence}
A set $\left\{\vct{s}_1,\ldots,\vct{s}_r \right\} \subseteq \left\{ \pm 1 \right\}^n$ of sign vectors  
is \emph{Schur independent} when the set
\begin{displaymath}
\{ \mathbf{e} \} \cup \{ \vct{s}_i \odot \vct{s}_j : 1 \leq i < j \leq r \} \subseteq \R^n
\quad\text{is linearly independent}.
\end{displaymath}
By extension, we also say that the sign matrix $\mtx{S} = \begin{bmatrix} \vct{s}_1 & \dots & \vct{s}_r \end{bmatrix} \in \{ \pm 1 \}^{n \times r}$ is Schur independent when its columns form a Schur independent set.
\end{definition}

Let us summarize the basic properties of Schur independent
sets~\cite{LP96:Facial-Structure,Tro18:Simplicial-Faces,KT19:Binary-Factorization-I}.

\begin{fact}[Schur independence] \label{fact:kt-schur}
Assume that the set $\mathcal{S} = \{ \vct{s}_1, \dots, \vct{s}_r \} \subseteq \{ \pm 1\}^n$
of sign vectors is Schur independent.  We have the following consequences.

\begin{enumerate}
\item	\label{it:schur->linear}
The family $\mathcal{S}$ is linearly independent.

\item	Each subset of $\mathcal{S}$ is Schur independent.

\item	For any choice $\vct{\xi} \in \{\pm 1\}^{r}$ of signs, the set
$\{ \xi_1 \vct{s}_1, \dots, \xi_r \vct{s}_r \}$ remains Schur independent.

\item	\label{it:schur-card}
The cardinality $r$ of the set $\mathcal{S}$ satisfies $r \leq \tfrac{1}{2} (1 + \sqrt{8n - 7})$.

\item	We can determine whether or not $\mathcal{S}$ is Schur independent
in polynomial time.
\end{enumerate}
\end{fact}

Schur independence is best understood as a kind of ``general position'' property
for sign vectors.  Roughly speaking, almost all collections of sign vectors are
Schur independent, provided that the cardinality $r$ meets the bound
stated in Fact~\ref{fact:kt-schur}\eqref{it:schur-card}.  This intuition
is quantified in the paper~\cite{Tro18:Simplicial-Faces}.

\subsubsection{Computation}

The main result of this paper is an algorithm for computing the
minimal asymmetric sign component decomposition of a low-rank matrix.
This algorithm succeeds precisely when the sign component is Schur independent.
Moreover, this condition is sufficient to ensure that the sign component
decomposition is essentially unique.

\begin{bigthm}[Sign component decomposition] \label{thm:scd-main}
Let $\mtx{B} \in \R^{n \times m}$ be a matrix that admits
a sign component decomposition $\mtx{B} = \mtx{SW}^\transp$
where
\begin{enumerate}
\item	The sign matrix $\mtx{S} \in \{ \pm 1 \}^{n \times r}$ is Schur independent;
\item	The weight matrix $\mtx{W} \in \R^{m \times r}$ has full column rank.
\end{enumerate}
Then the minimal sign component decomposition (with inner dimension $r$)
is determined up to simultaneous sign flips and permutations of the columns of the factors.
Algorithm~\ref{alg:asymfactor} computes this decomposition in time polynomial in $n + m$.
\end{bigthm}

\noindent
The uniqueness claim is a consequence of Theorem~\ref{thm:uniqueness},
while the computational claim follows from Theorem~\ref{thm:correct}.

Theorem~\ref{thm:scd-main} identifies a rich set of factorizable matrices
for which exact identification is always tractable and essentially unique. 
Moreover, existing denoising techniques allow us to compute the factorization
in the presence of gross errors; see Section~\ref{sec:robustness}.
Small perturbations appear more challenging;
we will study this problem in future work.

It is surprising that the exact sign component decomposition is tractable.
Most existing approaches to structured matrix factorization only produce
approximations, and many of these approaches lack rigorous guarantees.
The companion paper~\cite[Sec.~8]{KT19:Binary-Factorization-I} contains
a discussion of the related work.

\subsection{Binary component decomposition}
\label{sec:bcd-intro}

The asymmetric sign component decomposition also serves as a primitive
that allows us to compute other discrete matrix factorizations.
In this section, we turn to the problem of producing a decomposition
where one component takes values in the set $\{0, 1\}$.

\subsubsection{The decomposition}

Suppose that $\mtx{C} \in \R^{n \times m}$ is a rectangular matrix.  We consider
a decomposition of the form
\begin{equation} \label{eqn:bcd-matrix}
\mtx{C} = \mtx{ZW}^\transp
\quad\text{where}\quad
\mtx{Z} \in \{0, 1\}^{n \times r}
\quad\text{and}\quad
\mtx{W} \in \R^{m \times r}.
\end{equation}
The vector formulation of this decomposition is
\begin{equation} \label{eqn:bcd-vector}
\mtx{C} = \sum_{i=1}^r \vct{z}_i \vct{w}_i^\transp
\quad\text{where}\quad
\vct{z}_i \in \{0, 1\}^n
\quad\text{and}\quad
\vct{w}_i \in \R^m.
\end{equation}
We refer to~\eqref{eqn:bcd-matrix}--\eqref{eqn:bcd-vector} as
an (asymmetric) \emph{binary component decomposition} of the matrix $\mtx{C}$.
The left factor $\mtx{Z}$ is called the \emph{binary component},
and its columns $\vct{z}_i$ are also called \emph{binary components}.
The right factor $\mtx{W}$ is unconstrained;
we refer to it as a \emph{weight} matrix.

Every $n \times m$ matrix $\mtx{C}$ admits a superabundance of distinct
binary component decompositions~\eqref{eqn:bcd-matrix} where the inner dimension $r = n$.
We focus on the case where the matrix $\mtx{C}$ has low rank,
and the factorization is \emph{minimal}; that is, the inner
dimension $r$ in~\eqref{eqn:bcd-matrix} equals the rank of $\mtx{C}$.

\subsubsection{Schur independence}

We can reduce the problem of computing a binary component decomposition
to the problem of computing a sign component decomposition.

To do so, we first observe that there is an affine map that places
the binary vectors and sign vectors in one-to-one correspondence:
\begin{equation} \label{eqn:affine-map}
\mtx{F} : \{ 0, 1 \}^n \to \{ \pm 1 \}^n
\quad\text{where}\quad
\mtx{F} : \vct{z} \mapsto 2\vct{z} - \mathbf{e}
\quad\text{and}\quad
\mtx{F}^{-1} : \vct{s} \mapsto \tfrac{1}{2} (\vct{s} + \mathbf{e}).
\end{equation}
We can extend the map $\mtx{F}$ to a matrix by applying it to each column. 
This correspondence suggests that there should also be
a concept of Schur independence for binary vectors.
Here is the notion that suits our purposes.

\begin{definition}[Schur independence of binary vectors] \label{def:schur-independence-01}
A set $\{\vct{z}_1, \dots, \vct{z}_r \} \subseteq \{0,1\}^n$
of binary vectors is \emph{Schur independent} when the set
\begin{displaymath}
\{ \mathbf{e} \} \cup \{ \vct{z}_i : 1 \leq i \leq r \} \cup
\{ \vct{z}_i \odot \vct{z}_j : 1 \leq i < j \leq r \} \subseteq \R^n
\quad \text{is linearly independent.}
\end{displaymath}
By extension, we say that a binary matrix
$\mtx{Z} = \begin{bmatrix} \vct{z}_1 & \dots & \vct{z}_r \end{bmatrix} \in \{0, 1\}^{n \times r}$
is Schur independent when its columns compose a Schur independent set.
\end{definition}

The following result~\cite[Prop.~6.3]{KT19:Binary-Factorization-I}
describes the precise connection between the two flavors of Schur independence.

\begin{fact}[Kueng \& Tropp] \label{fact:bcd-scd}
The binary matrix $\mtx{Z} \in \{0,1\}^{r}$ is Schur independent
if and only if the sign matrix $\begin{bmatrix} \mtx{F}(\mtx{Z}) & \mathbf{e} \end{bmatrix} \in \{\pm 1\}^{r+1}$
is Schur independent.
\end{fact}

\subsubsection{Computation}

With these definitions at hand, we can state our main result on
binary component decompositions.

\begin{bigthm}[Binary component decomposition] \label{thm:bcd-main}
Let $\mtx{C} \in \R^{n \times m}$ be a matrix that admits
a binary component decomposition $\mtx{C} = \mtx{ZW}^\transp$ where
\begin{enumerate}
\item	The binary matrix $\mtx{Z} \in \{ 0, 1 \}^{n \times r}$ is Schur independent;
\item	The weight matrix $\mtx{W} \in \R^{m \times r}$ has full column rank.
\end{enumerate}
Then the minimal binary component decomposition (with inner dimension $r$)
is determined up to simultaneous permutation of the columns of the factors.
Algorithm~\ref{alg:binfactor} computes the decomposition in time polynomial in $n + m$.
\end{bigthm}

\noindent
The uniqueness claim is established in Theorem~\ref{thm:bin-uniqueness},
and the computational claim appears in Theorem~\ref{thm:bcd-correct}.

\begin{algorithm}[t]
\begin{algorithmic}[1]
\caption{\textit{Asymmetric sign component decomposition~\eqref{eqn:scd-matrix}--\eqref{eqn:scd-vector} of a matrix with a Schur independent sign component}.  Implements the procedure from Section~\ref{sec:compute}.}
\label{alg:asymfactor}

\Require	Rank-$r$ matrix $\mtx{B} \in \R^{n \times m}$ that satisfies the conditions of Theorem~\ref{thm:scd-main}.
\Ensure		Sign matrix $\tilde{\mtx{S}} \in \{ \pm 1 \}^{n \times r}$ and weight matrix $\tilde{\mtx{W}} \in \R^{m \times r}$
where $\mtx{B} = \tilde{\mtx{S}} \tilde{\mtx{W}}^\transp$.
\Statex
\Function{AsymSignComponentDecomposition}{$\mtx{B}$}
\State	$[n, m] \gets \texttt{size}(\mtx{B})$ and $r \gets \rank(\mtx{A})$
\State	$\mtx{U} \gets \texttt{orth}(\mtx{A})$
	\Comment Find a basis for the range of $\mtx{A}$
\State	Find the solution $(\mtx{X}_{\star}, \mtx{Y}_{\star})$ to the semidefinite program
\begin{displaymath}
\begin{aligned}
&\underset{\mtx{X}\in \Sym_n, \mtx{Y} \in \Sym_m}{\minimize}\quad &&\trace(\mtx{Y}) \\
	&\subjto\quad
	&&\text{$\trace(\mtx{U}^\transp \mtx{X} \mtx{U}) = n$ and $\diag(\mtx{X}) = \mathbf{e}$;} \\
	&&&\begin{bmatrix} \mtx{X} & \mtx{B} \\ \mtx{B}^\transp & \mtx{Y} \end{bmatrix} \psdge \mtx{0} 
\end{aligned}
\end{displaymath}

\State	Apply Algorithm~\ref{alg:symfactor} to $\mtx{X}_{\star}$
to obtain a symmetric sign component decomposition~\eqref{eqn:sym-scd-matrix}:
\begin{displaymath}
\mtx{X}_{\star} = \tilde{\mtx{S}} \, \diag(\tilde{\vct{\tau}}) \, \tilde{\mtx{S}}^\transp
\quad\text{where $\tilde{\mtx{S}} \in \{\pm 1\}^{n \times r}$}
\end{displaymath}

\State	Find the solution $\tilde{\mtx{W}} \in \R^{m \times r}$ to the linear system
$\mtx{B} = \tilde{\mtx{S}} \tilde{\mtx{W}}^\transp$
\EndFunction
\end{algorithmic}
\end{algorithm}

\begin{algorithm}[t]
\begin{algorithmic}[1]
\caption{\textit{Binary component decomposition~\eqref{eqn:bcd-matrix} of a matrix with a Schur independent binary component.} Implements the procedure from Section~\ref{sec:bcd-compute}.}
\label{alg:binfactor}

\Require	Rank-$r$ matrix $\mtx{C} \in \Sym_n$ that satisfies the conditions of Theorem~\ref{thm:bcd-main}
\Ensure		Binary matrix $\tilde{\mtx{Z}} \in \{0,1\}^{n \times r}$
and weight matrix $\tilde{\mtx{W}}_{+} \in \R^{m \times r}$ for which $\mtx{C} = \tilde{\mtx{Z}} \tilde{\mtx{W}}_+^\transp$.

\Statex
\Function{AsymBinaryComponentDecomposition}{$\mtx{C}$}

\State	Form the matrix $\mtx{B} = 2 \mtx{C} - \mathbf{E}$

\State	Apply Algorithm~\ref{alg:asymfactor} to $\mtx{B}$ to obtain
a sign component decomposition
\begin{displaymath}
\mtx{B} = \tilde{\mtx{S}} \tilde{\mtx{W}}^\transp
	= \begin{bmatrix} \tilde{\vct{s}}_1 & \dots & \tilde{\vct{s}}_{r+1} \end{bmatrix}
	\begin{bmatrix} \tilde{\vct{w}}_1 & \dots & \tilde{\vct{w}}_{r+1} \end{bmatrix}^\transp
\end{displaymath}

\State	Find the index $i$ and sign $\varphi \in \{\pm 1\}$
where $\tilde{\vct{s}}_{i} = \varphi \mathbf{e}$.  Permute the sequences to interchange
\begin{displaymath}
\tilde{\vct{s}}_i \leftrightarrow \tilde{\vct{s}}_{r+1}
\quad\text{and}\quad
\tilde{\vct{w}}_i \leftrightarrow \tilde{\vct{w}}_{r+1}
\end{displaymath}

\State	Find the solution $\vct{\xi} \in \R^{r}$ to the linear system
\begin{displaymath}
\begin{bmatrix} \tilde{\vct{w}}_1 & \dots & \tilde{\vct{w}}_r \end{bmatrix} \vct{\xi}
	= \varphi \tilde{\vct{w}}_{r+1} + \mathbf{e}
\end{displaymath}

\State	Set $\tilde{\vct{z}}_i = \tfrac{1}{2} (\xi_i \tilde{\vct{s}}_i + \mathbf{e})$
for each index $i = 1, \dots, r$

\State	Define the binary matrix $\tilde{\mtx{Z}} = \begin{bmatrix} \tilde{\vct{z}}_1 & \dots & \tilde{\vct{z}}_r \end{bmatrix}$
and the weight matrix $\tilde{\mtx{W}}_{+} = \begin{bmatrix} \xi_1 \tilde{\vct{w}}_1 & \dots & \xi_r \tilde{\vct{w}}_r \end{bmatrix}$
\EndFunction
\end{algorithmic}
\end{algorithm}

\subsection{The planted sign basis problem}

Theorem~\ref{thm:scd-main} and Theorem~\ref{thm:bcd-main}
allow us to solve some interesting combinatorial problems
in linear algebra.

\begin{problem}[Planted sign basis] \label{prob:planted}
Let $\mathsf{L} \subseteq \R^n$ be an $r$-dimensional subspace that admits a sign basis:
\begin{displaymath}
\mathsf{L} = \lspan\{ \vct{s}_1, \dots, \vct{s}_r \}
\quad\text{where each $\vct{s}_i \in \{ \pm 1 \}^n$.}
\end{displaymath}
Given the subspace $\mathsf{L}$, find a sign basis for the subspace.
\end{problem}

To clarify, we can assume that the problem data is a matrix $\mtx{B} \in \R^{n \times m}$
whose range equals the $r$-dimensional subspace $\mathsf{L}$.
We must output a set of $r$ sign vectors that generates the subspace.
The brute force approach may require us to sift through around
$2^{nr}$
families of sign vectors.  Is it possible to solve
the problem more efficiently?

Let us outline a solution for Problem~\ref{prob:planted}
in the case where $\mathsf{L}$ has a sign basis $\{ \vct{s}_1, \dots, \vct{s}_r \} \subseteq \{\pm 1\}^n$
that is Schur independent.  This is a rather mild deterministic condition,
provided that the dimension $r$ of the subspace satisfies $r < \tfrac{1}{2} (1 + \sqrt{8n - 7})$.
The hypothesis also guarantees that the basis is determined up to permutation and sign flips,
per Theorem~\ref{thm:scd-main}.

Here is how we solve the problem.
Let $\mtx{B} \in \R^{n \times m}$ be a matrix whose range coincides with the subspace $\mathsf{L}$.
A Schur independent set is linearly independent,
so we can write the matrix in the form $\mtx{B} = \mtx{SW}^\transp$, where
$\mtx{S} = \begin{bmatrix} \vct{s}_1 & \dots & \vct{s}_r \end{bmatrix} \in \{ \pm 1 \}^{n\times r}$
and the weight matrix $\mtx{W} \in \R^{m \times r}$ has full column rank.
As a consequence, we can apply Algorithm~\ref{alg:asymfactor} to the matrix $\mtx{B}$
to obtain a sign component decomposition $\mtx{B} = \tilde{\mtx{S}} \tilde{\mtx{W}}^\transp$.
Theorem~\ref{thm:scd-main} ensures that the columns of $\tilde{\mtx{S}}$
coincide with the columns of $\mtx{S}$ up to sign flips and permutations.
In other words, the columns of $\tilde{\mtx{S}}$ compose the (unique)
sign basis that generates $\mathsf{L}$.
In summary, we can solve Problem~\ref{prob:planted} for any subspace that
is spanned by a Schur independent family of sign vectors.

A similar procedure, using Algorithm~\ref{alg:binfactor}, allows us
to solve a variant of Problem~\ref{prob:planted} where we seek
a planted \emph{binary} basis for a subspace.  Indeed, if
a subspace is generated by a Schur independent family of binary
vectors, then we can identify the basis up to permutation.

\subsection{Roadmap}
\label{sec:roadmap}

We continue with a discussion about symmetric sign component decompositions
in Section~\ref{sec:sym-scd}.
In Section~\ref{sec:exist-unique},
we develop basic results about existence and uniqueness of
asymmetric sign component decompositions.
Section~\ref{sec:compute} explains how to
compute an sign component decomposition.
We turn to binary component decomposition in Section~\ref{sec:bcd-full}.
Finally, in Section~\ref{sec:robustness}, 
we state some results on robustness of sign component
decomposition which we prove in the appendices.
For a discussion of related work,
see the companion paper~\cite[Sec.~8]{KT19:Binary-Factorization-I}.

\section{Symmetric sign component decomposition}
\label{sec:sym-scd}

This section contains a summary of the principal results
from the companion paper~\cite{KT19:Binary-Factorization-I}.
These results play a core role in our study of
asymmetric factorizations.

\subsection{Signed permutations}

Matrix factorizations are usually not fully determined because they
are invariant under some group of symmetries.  For example, consider
the decomposition of a psd matrix as the outer product of two symmetric factors:
\begin{displaymath}
\mtx{A} = \mtx{BB}^\transp = (\mtx{BQ})(\mtx{BQ})^{\transp}
\quad\text{for each orthogonal $\mtx{Q}$.}
\end{displaymath}
Each of the factorizations on the right is equally valid,
because there is no constraint that forbids rotations.

For binary component decompositions,
permutations compose the relevant symmetry group.

\begin{definition}[Permutation]
A \emph{permutation} on $r$ letters is an element $\pi$ of the
symmetric group $\mathsf{Sym}_r$.  A permutation $\pi$ acts
on $\R^r$ via the linear map $\vct{x} \mapsto (x_{\pi(1)}, \dots, x_{\pi(r)})$.
This linear map can be represented by the \emph{permutation matrix}
$\mtx{\Pi} \in \R^{r \times r}$ whose entries take the form
$(\mtx{\Pi})_{ij} = 1$ where $j = \pi(i)$ and are zero otherwise.
A permutation matrix is orthogonal: $\mtx{\Pi} \mtx{\Pi}^\transp = \Id = \mtx{\Pi}^\transp \mtx{\Pi}$.
\end{definition}

For sign component decompositions,
the signed permutations make up the relevant symmetry group.

\begin{definition}[Signed permutation]
A \emph{signed permutation} on $r$ letters is a pair $(\pi, \vct{\xi}) \in \mathsf{Sym}_r \times \{ \pm 1 \}^r$
consisting of a permutation $\pi$ on $r$ letters and a sign vector $\vct{\xi} \in \left\{ \pm 1 \right\}^r$.
The signed permutation $(\pi, \vct{\xi})$ acts on $\R^r$ via the
linear map $\vct{x} \mapsto (\xi_1 x_{\pi(1)}, \dots, \xi_r x_{\pi(r)})$.
This linear map can also be represented by the \emph{signed permutation matrix} $\mtx{\Pi} \in \R^{r \times r}$
whose entries satisfy $(\mtx{\Pi})_{ij} = \xi_i$ when $j = \pi(i)$ and are otherwise zero.
Each signed permutation matrix is orthogonal. 
\end{definition}

\subsection{Symmetric sign component decomposition}

In the companion paper~\cite{KT19:Binary-Factorization-I},
we explored the problem of computing a (symmetric) sign component
decomposition of a correlation matrix.
This research provides the foundation for the
asymmetric sign component decomposition.  Let us take a moment
to present the principal definitions and results from the associated work.

Let $\mtx{A} \in \Sym_n$ be a correlation matrix; that is,
$\mtx{A}$ is psd with all diagonal entries equal to one.
We say that $\mtx{A}$ has a \emph{symmetric sign component decomposition}
when
\begin{equation} \label{eqn:sym-scd-matrix}
\mtx{A} = \mtx{S} \, \diag(\vct{\tau}) \, \mtx{S}^\transp
\quad\text{where}\quad
\text{$\mtx{S} \in \{ \pm 1 \}^{n \times r}$ and $\vct{\tau} \in \Delta_r^+$.}
\end{equation}
In vector form,
\begin{displaymath}
\mtx{A} = \sum_{i=1}^r \tau_i \vct{s}_i \vct{s}_i^\transp
\quad\text{where}\quad
\vct{s}_i \in \{ \pm 1 \}^n
\quad\text{and}\quad
(\tau_1, \dots, \tau_r) \in \Delta_r^+.
\end{displaymath}
The sign matrix $\mtx{S}$ is called the \emph{sign component},
while the positive diagonal matrix, $\diag(\vct{\tau})$,
is a list of convex coefficients.
Not all correlation matrices admit a symmetric sign component decomposition,
nor does the factorization need to be uniquely determined;
see~\cite{KT19:Binary-Factorization-I} for a full discussion.

The situation improves markedly when the sign component $\mtx{S}$ is Schur independent.
In this case, the sign component decomposition is essentially unique,
and we can compute it by means of an efficient algorithm~\cite[Thm.~I]{KT19:Binary-Factorization-I}.

\begin{fact}[Kueng \& Tropp] \label{fact:kt-scd}
Let $\mtx{A} \in \Sym_n$ be a correlation matrix that admits
a sign component decomposition: 
\begin{displaymath} 
\mtx{A} = \mtx{S} \, \diag(\vct{\tau}) \, \mtx{S}^\transp
\quad\text{where}\quad
\text{$\mtx{S} \in \{\pm 1\}^{n \times r}$ is Schur independent and $\vct{\tau} \in \Delta_r^+$.}
\end{displaymath}
Then the sign component decomposition of $\mtx{A}$ is determined up to
signed permutation.  Moreover,
with probability one, Algorithm~\ref{alg:symfactor}
computes the sign component decomposition.
That is, the output is a pair $(\tilde{\mtx{S}}, \tilde{\vct{\tau}})$
where the sign matrix $\tilde{\mtx{S}} = \mtx{S \Pi}$
and the convex coefficients $\tilde{\vct{\tau}}_i = | ( \mtx{\Pi}^\transp \vct{\tau} )_i |$ ($1 \leq i \leq r$),
for a signed permutation matrix $\mtx{\Pi} \in \R^{r \times r}$.
\end{fact}

\begin{algorithm}[t]
\begin{algorithmic}[1]
\caption{\textit{Symmetric sign component decomposition~\eqref{eqn:sym-scd-matrix} of a correlation matrix with a Schur independent sign component.}  Duplicates~\cite[Alg.~1]{KT19:Binary-Factorization-I}.}
\label{alg:symfactor}

\Require	Rank-$r$ correlation matrix $\mtx{A} \in \Sym_n$ that satisfies the conditions of Fact~\ref{fact:kt-scd}.
\Ensure		Sign matrix $\tilde{\mtx{S}} \in \{ \pm 1 \}^{n \times r}$ and convex coefficients $\tilde{\vct{\tau}} \in \Delta_r^+$ where $\mtx{A} = \tilde{\mtx{S}} \, \diag(\tilde{\vct{\tau}}) \, \tilde{\mtx{S}}^\transp$.
\Statex
\Function{SignComponentDecomposition}{$\mtx{A}$}

\State	$[n, \sim] \gets \texttt{size}(\mtx{A})$ and $r \gets \rank(\mtx{A})$

\For{$i = 1$ to $(r-1)$}
\State	$\mtx{U} \gets \texttt{orth}(\mtx{A})$
	\Comment Find a basis for the range of $\mtx{A}$
\State	$\vct{g} \gets \texttt{randn}(n, 1)$
	\Comment Draw a random direction
\State	Find the solution $\mtx{X}_{\star}$ to the semidefinite program
\begin{displaymath}
\underset{\mtx{X} \in \mathbb{S}^n}{\maximize} \quad \vct{g}^\transp \mtx{X} \vct{g}
\quad\subjto\quad \text{$\trace \left( \mtx{U}^\transp \mtx{X} \mtx{U} \right) = n$ and
$\diag(\mtx{X}) = \mathbf{e}$ and $\mtx{X} \psdge \mtx{0}$}
\end{displaymath}
\State	Factorize the rank-one matrix $\mtx{X}_{\star} = \tilde{\vct{s}}_i \tilde{\vct{s}}_i^\transp$
\Comment	Extract a sign component
\State	Find the solution $\zeta_{\star}$ to the semidefinite program 
\begin{align*}
\underset{\zeta \in \mathbb{R}}{\maximize} \quad \zeta \quad \subjto \quad
\zeta \mtx{A} + (1-\zeta) \mtx{X}_{\star} \psdge \mtx{0}
\end{align*}
\State	$\mtx{A} \gets \zeta_{\star} \mtx{A} + (1-\zeta_{\star}) \mtx{X}_{\star}$
\EndFor
\State	Factorize the rank-one matrix $\mtx{A} = \tilde{\vct{s}}_r\tilde{\vct{s}}_r^\transp$
\Comment	$\rank(\mtx{A}) = 1$ in final iteration
\State	Define the matrix $\tilde{\mtx{S}} = \begin{bmatrix} \tilde{\vct{s}}_1 & \dots & \tilde{\vct{s}}_r \end{bmatrix}$,
and find the solution $\tilde{\vct{\tau}} \in \Delta_r^+$ to the linear system
\begin{displaymath}
\mtx{A} = \tilde{\mtx{S}} \, \diag(\tilde{\vct{\tau}}) \, \tilde{\mtx{S}}^\transp
\end{displaymath}
\EndFunction
\end{algorithmic}
\end{algorithm}

A major ingredient in the proof of Fact~\ref{fact:kt-scd} is a characterization
of the set of correlation matrices that are generated by a Schur independent
family of sign vectors~\cite[Thm.~3.6]{KT19:Binary-Factorization-I}.

\begin{fact}[Kueng \& Tropp] \label{fact:kt-exposed}
Suppose that $\mtx{S} \in \{ \pm 1 \}^{n \times r}$ is a Schur independent sign matrix,
and let $\mtx{P} \in \Sym_n$ be the orthogonal projector onto $\range(\mtx{S})$.  Then
\begin{equation} \label{eqn:Scorr-Xcorr}
\{ \mtx{S} \, \diag(\vct{\tau}) \, \mtx{S}^\transp : \vct{\tau} \in \Delta_r \}
= \{ \mtx{X} \in \Sym_n : \text{$\trace(\mtx{PX}) = n$ {\rm and} $\diag(\mtx{X}) = \mathbf{e}$ {\rm and} $\mtx{X} \psdge \mtx{0}$} \}.
\end{equation}
\end{fact}

Fact~\ref{fact:kt-exposed} is a powerful tool for working with
sign component decompositions.  Indeed, we can compute the
projector $\mtx{P}$ onto the range of a Schur independent sign
matrix $\mtx{S}$ directly from any particular correlation matrix
$\mtx{A} = \mtx{S} \, \diag(\vct{\tau}) \, \mtx{S}^\transp$ with $\vct{\tau} \in \Delta_r^+$.
As a consequence, the identity~\eqref{eqn:Scorr-Xcorr} provides an alternative
representation for the set of all correlation matrices with sign
component $\mtx{S}$, which allows us to optimize over this set.
Fact~\ref{fact:kt-exposed} also plays a critical role in our method
for computing an asymmetric sign component decomposition.

\section{Existence and uniqueness of the asymmetric sign component decomposition}
\label{sec:exist-unique}

In this section, we begin our investigation of the asymmetric sign
component decomposition.  We lay out some of the basic questions,
and we start to deliver the answers.

\subsection{Questions}

This paper addresses four fundamental problems raised by the
definition~\eqref{eqn:scd-matrix}--\eqref{eqn:scd-vector}
of the asymmetric sign component decomposition:
\begin{enumerate}
\item \textbf{Existence:} Which matrices admit a sign component decomposition?
\item \textbf{Uniqueness:} When is the sign component decomposition unique, modulo symmetries?
\item \textbf{Computation:} How can we find a sign component decomposition in polynomial time? 
\item \textbf{Robustness:} How can we find a sign component decomposition from a noisy observation?
\end{enumerate}

This section treats the structural questions about existence and
uniqueness of the sign component decomposition, and
Section~\ref{sec:compute} explains how we can compute the factorization.
Last, Section~\ref{sec:robustness} describes some situations where
we can extract a sign component decomposition from imperfect data.

\subsection{Existence}

We quickly dispatch the first question, which concerns the existence of
asymmetric sign component decompositions.  

\begin{proposition}[Sign component decomposition: Existence]
\label{prop:existence}
Every matrix $\mtx{B} \in \R^{n \times m}$ admits a
sign component decomposition~\eqref{eqn:scd-matrix}
with inner dimension $r = n$.
\end{proposition}

\begin{proof}
Let $\mtx{S} \in \{\pm 1\}^{n \times n}$ be a nonsingular matrix of signs.
Define the second factor $\mtx{W}^\transp = \mtx{S}^{\dagger} \mtx{B}$.
\end{proof}

As an aside, we remark that nonsingular sign matrices are ubiquitous.
Indeed, a uniformly random element of $\{\pm 1\}^{n \times n}$ is nonsingular
with exceedingly high probability~\cite{Tik18:Singularity-Bernoulli}. 

Proposition~\ref{prop:existence} ensures that every matrix
has an exorbitant number of sign component decompositions.
Therefore, we need to burden the factorization with extra
conditions before it is determined uniquely.
We intend to focus on minimal factorizations, where the target matrix $\mtx{B} \in \R^{n \times m}$
has rank $r < n$, and the number $r$ of sign components coincides with the rank.

\subsection{Symmetries}

Like many other matrix factorizations, the sign component decomposition
has some symmetries that we can never resolve.  Before we can
turn to the question of uniqueness, we need to discuss
invariants of the factorization.

Signed permutations preserve the sign component decomposition~\eqref{eqn:scd-matrix}--\eqref{eqn:scd-vector}
in the following sense.  Suppose that $\mtx{B} \in \R^{n \times m}$ has the sign component decomposition
\begin{displaymath}
\mtx{B} = \mtx{SW}^\transp =
	\begin{bmatrix} \vct{s}_1 & \dots & \vct{s}_r \end{bmatrix}
	\begin{bmatrix} \vct{w}_1 & \dots & \vct{w}_r \end{bmatrix}^\transp =
	\sum_{i=1}^r \vct{s}_i \vct{w}_i^\transp.
\end{displaymath}
For a signed permutation $(\pi, \vct{\xi})$ on $r$ letters
with associated signed permutation matrix $\mtx{\Pi}$,
we have
\begin{displaymath}
\begin{aligned}
\mtx{B} = \mtx{S} \mtx{\Pi} \mtx{\Pi}^{-1} \mtx{W}^\transp
	= (\mtx{S\Pi})(\mtx{W}\mtx{\Pi})^\transp
	&= \begin{bmatrix} \xi_1 \vct{s}_{\pi(1)} & \dots & \xi_r \vct{s}_{\pi(r)} \end{bmatrix}
	\begin{bmatrix} \xi_1 \vct{w}_{\pi(1)} & \dots & \xi_r \vct{w}_{\pi(r)} \end{bmatrix}^\transp. 
\end{aligned}
\end{displaymath}
Observe that $\mtx{S\Pi} \in \{ \pm 1 \}^{n \times r}$ remains a sign matrix.
Therefore, $\mtx{SW}^\transp$ and $(\mtx{S\Pi})(\mtx{W\Pi})^\transp$
are both sign component decompositions of $\mtx{B}$.

We have no cause to prefer one of the sign component decompositions
induced by a signed permutation over the others.  Thus, it is appropriate
to treat them all as equivalent.

\begin{definition}[Sign component decomposition: Equivalence]
Suppose that $\mtx{B} = \mtx{SW}^\transp$ and $\mtx{B} = \tilde{\mtx{S}} \tilde{\mtx{W}}^\transp$
are two sign component decompositions~\eqref{eqn:scd-matrix} with the same inner dimension $r$.
We say that the decompositions are \emph{equivalent}
if there is a signed permutation matrix $\mtx{\Pi}$ for which $\tilde{\mtx{S}} = \mtx{S} \mtx{\Pi}$
and $\tilde{\mtx{W}} = \mtx{W} \mtx{\Pi}$.

Alternatively, consider two sign component decompositions 
$\mtx{B} = \sum_{i=1}^r \vct{s}_i \vct{w}_i^\transp$ and $\mtx{B} = \sum_{i=1}^r \tilde{\vct{s}}_i \tilde{\vct{w}}_i$
with the same number $r$ of terms.  The decompositions are \emph{equivalent}
if there is a signed permutation $(\pi, \vct{\xi})$ on $r$ letters for which
$\tilde{\vct{s}}_i = \xi_i \vct{s}_{\pi(i)}$ and $\tilde{\vct{w}}_i = \xi_i \vct{w}_{\pi(i)}$
for each $i = 1, \dots, r$.
\end{definition}

\subsection{The role of Schur independence}

As we have just seen, signed permutations preserve the class of sign component
decompositions of a given matrix.  Meanwhile, the proof of
Proposition~\ref{prop:existence} warns us that we can sometimes
map one sign component decomposition to an inequivalent decomposition
via an invertible transformation.
Remarkably, we can preclude the latter phenomenon by narrowing our attention to
Schur independent sign matrices.  In this case, sign permutations are
the \emph{only} invertible transformations that respect the sign structure.

\begin{proposition}[Schur independence: Transformations] \label{prop:schur-transform}
Let $\mtx{S} \in \{ \pm 1\}^{n \times r}$
be a Schur independent sign matrix, and let $\mtx{Q} \in \R^{r \times r}$ be an invertible matrix.
Then $\mtx{SQ} \in \{ \pm 1 \}^{n \times r}$ is a sign matrix if and only if
$\mtx{Q}$ is a signed permutation.
\end{proposition}

\begin{proof}
If $\mtx{Q}$ is a signed permutation, then it is immediate that $\mtx{SQ}$ is a sign matrix.
The reverse implication is the more interesting fact.

Introduce notation for the columns of the matrices under discussion:
\begin{displaymath}
\mtx{S} = \begin{bmatrix} \vct{s}_1 & \dots & \vct{s}_r \end{bmatrix} \in \{ \pm 1 \}^{n \times r}
\quad\text{and}\quad
\mtx{Q} = \begin{bmatrix} \vct{q}_1 & \dots & \vct{q}_r \end{bmatrix} \in \R^{r\times r}
\quad\text{and}\quad
\mtx{SQ} = \begin{bmatrix} \tilde{\vct{s}}_1 & \dots & \tilde{\vct{s}}_r \end{bmatrix} \in \{ \pm 1 \}^{n \times r}.
\end{displaymath}
For each index $1 \leq k \leq r$, the $k$th column $\tilde{\vct{s}}_k$
of the matrix $\mtx{SQ}$ satisfies
\begin{displaymath}
\tilde{\vct{s}}_k = \mtx{S} \vct{q}_k = \sum_{i=1}^r \ip{ \mathbf{e}_i }{ \vct{q}_k }\, \vct{s}_i.
\end{displaymath}
By assumption, $\tilde{\vct{s}}_k$ is a sign vector, so
\begin{displaymath}
\mathbf{e} = \tilde{\vct{s}}_k \odot \tilde{\vct{s}}_k
	= \sum_{i,j=1}^r \ip{ \mathbf{e}_i }{ \vct{q}_k } \ip{ \mathbf{e}_j }{ \vct{q}_k }( \vct{s}_i \odot \vct{s}_j )
	= \left( \sum_{i=1}^r \ip{ \mathbf{e}_i }{ \vct{q}_k }^2 \right) \mathbf{e}
	+ 2 \sum_{i < j} \ip{ \mathbf{e}_i }{ \vct{q}_k } \ip{ \mathbf{e}_j }{ \vct{q}_k }( \vct{s}_i \odot \vct{s}_j ).
\end{displaymath}
Schur independence of the matrix $\mtx{S}$ ensures that the family
$\{ \mathbf{e} \} \cup \{ \vct{s}_i \odot \vct{s}_j : i < j \} \subset \R^r$
is linearly independent.  As a consequence,
\begin{displaymath}
\sum_{i=1}^r \ip{ \mathbf{e}_i }{ \vct{q}_k }^2 = 1
\quad\text{and}\quad
\ip{ \mathbf{e}_i }{ \vct{q}_k } \ip{ \mathbf{e}_j }{ \vct{q}_k } = 0
\quad\text{when $i \neq j$.}
\end{displaymath}
Since $\vct{q}_k$ solves this quadratic system, it must be a signed standard basis vector:
$\vct{q}_k = \xi_k \mathbf{e}_{\pi(k)} \in \R^r$ for a sign $\xi_k \in \{ \pm 1 \}$
and an index $\pi(k) \in \{ 1, \dots, r \}$.  Since the matrix $\mtx{Q}$ is invertible,
it must be the case that $\pi$ is a permutation on $r$ letters.  It follows  that
$\mtx{Q}$ is a signed permutation.
\end{proof}

\subsection{Uniqueness}

With this preparation, we can delineate circumstances where the (minimal)
sign component decomposition of a low-rank matrix is unique up to equivalence.

\begin{theorem}[Sign component decomposition: Uniqueness] \label{thm:uniqueness}
Consider a matrix $\mtx{B} \in \R^{n \times m}$
that admits a sign component decomposition $\mtx{B} = \mtx{SW}^\transp$.
Assume that
\begin{enumerate}
\item	The sign matrix $\mtx{S} \in \{ \pm 1 \}^{n \times r}$ is Schur independent;
\item	The weight matrix $\mtx{W} \in \R^{m \times r}$ has full column rank.
\end{enumerate}
Then all minimal sign component decompositions of $\mtx{B}$ (with inner dimension $r$) are equivalent.
\end{theorem}

\begin{proof}
The sign matrix $\mtx{S} \in \{ \pm 1 \}^{n \times r}$ has full column rank because it is Schur independent
(Fact~\ref{fact:kt-schur}\eqref{it:schur->linear}),
while the weight matrix $\mtx{W} \in \R^{m \times r}$ has full column rank by assumption.
We discover that the matrix $\mtx{B} = \mtx{SW}^\transp$ has rank $r$.  Therefore,
every sign component decomposition of $\mtx{B}$ has inner dimension at least $r$,
and the distinguished decomposition has the minimal inner dimension.

Suppose that $\mtx{B} = \tilde{\mtx{S}} \tilde{\mtx{W}}^\transp$
is another sign component decomposition with inner dimension $r$.  Since $\mtx{B}$ has rank $r$,
both factors $\tilde{\mtx{S}}$ and $\tilde{\mtx{W}}$ must have full column rank.  As a consequence,
there is an invertible transformation $\mtx{Q} \in \R^{r \times r}$ for which
$\tilde{\mtx{S}} = \mtx{SQ}$.  Since $\mtx{S}$ is a Schur independent sign matrix and $\tilde{\mtx{S}}$
is a sign matrix, Proposition~\ref{prop:schur-transform} forces $\mtx{Q}$ to be a signed permutation.
Now, we have the chain of identities
\begin{displaymath}
\mtx{SW}^\transp = \mtx{B} = \tilde{\mtx{S}}\tilde{\mtx{W}}^\transp = \mtx{SQ} \tilde{\mtx{W}}^{\transp}
	= \mtx{S} (\tilde{\mtx{W}} \mtx{Q}^\transp)^{\transp}.
\end{displaymath}
Since the matrix $\mtx{S}$ has full column rank, we can cancel $\mtx{S}$
to see that $\tilde{\mtx{W}} \mtx{Q}^\transp = \mtx{W}$.
The signed permutation $\mtx{Q}$ is orthogonal, so
it follows that $\tilde{\mtx{W}} = \mtx{W}\mtx{Q}$.

To summarize, we have been given two sign component decompositions
$\mtx{B} = \mtx{SW}^\transp = \tilde{\mtx{S}}\tilde{\mtx{W}}^\transp$
with inner dimension $r$.  We have shown that they are related by $\tilde{\mtx{S}} = \mtx{SQ}$
and $\tilde{\mtx{W}} = \mtx{WQ}$ for a signed permutation $\mtx{Q}$.
Therefore, the two decompositions are equivalent.
\end{proof}

Theorem~\ref{thm:uniqueness} describes conditions under which the minimal
sign component decomposition of a matrix is uniquely determined.
It is natural to demand that both the left and the right factors
have full column rank.  The geometry of the factorization problem
dictates the stronger requirement that the sign matrix is Schur independent.
As we have discussed, most families of $r < \tfrac{1}{2} (1 + \sqrt{8n - 7})$ sign
vectors are Schur independent, so this condition holds for a rich
class of matrices.

\section{Computation of the asymmetric sign component decomposition}
\label{sec:compute}

In this section, we derive and justify Algorithm~\ref{alg:symfactor},
which computes the asymmetric sign component decomposition of a matrix
whose sign component is Schur independent.  We establish the following result.

\begin{theorem}[Sign component decomposition: Computation] \label{thm:correct}
Consider a matrix $\mtx{B} \in \R^{n \times m}$ that admits
a sign component decomposition $\mtx{B} = \mtx{SW}^\transp$.
Assume that
\begin{enumerate}
\item	The sign matrix $\mtx{S} \in \{ \pm 1 \}^{n \times r}$ is Schur independent;
\item	The weight matrix $\mtx{W} \in \R^{m \times r}$ has full column rank.
\end{enumerate}
Then, with probability one, Algorithm~\ref{alg:asymfactor} identifies
the minimal sign component decomposition, up to signed permutation. 
That is, the output is a pair $(\tilde{\mtx{S}}, \tilde{\mtx{W}})
\in \{ \pm 1 \}^{n \times r} \times \R^{m \times r}$ where
$\tilde{\mtx{S}} = \mtx{S\Pi}$ and $\tilde{\mtx{W}} = \mtx{W\Pi}$
for a signed permutation $\mtx{\Pi} \in \R^{r \times r}$.
\end{theorem}

\noindent
We prove Theorem~\ref{thm:correct} below in Section~\ref{sec:correct-pf}.

\subsection{Factorization and semidefinite programming}

Although constrained matrix factorization is viewed as a challenging problem,
certain aspects are simpler than they appear.  In particular, we can expose
properties of the components of a matrix factorization
by means of a semidefinite constraint.

\begin{fact}[Factorization constraint] \label{fact:factorization}
Let $\mtx{B}\in \R^{n \times m}$ be a matrix.  The semidefinite relation
\begin{equation} \label{eqn:factor-sdr}
\begin{bmatrix} \mtx{X} & \mtx{B} \\ \mtx{B}^\transp & \mtx{Y} \end{bmatrix} \psdge \mtx{0}
\end{equation}
enforces a factorization of $\mtx{B}$ in the following sense.

\begin{enumerate}
\item	If $\mtx{B} = \mtx{UV}^\transp$, 
then~\eqref{eqn:factor-sdr} holds when $\mtx{X} = \mtx{UU}^\transp$ and $\mtx{Y} =\mtx{VV}^\transp$.

\item	If~\eqref{eqn:factor-sdr} holds, then we can decompose $\mtx{B} = \mtx{UV}^\transp$
into factors $\mtx{U} \in \R^{n \times r}$ and $\mtx{V} \in \R^{m \times r}$ that
satisfy $\mtx{X} = \mtx{UU}^\transp$ and $\mtx{Y} = \mtx{VV}^\transp$.
The inner dimension meets the bound $r \leq m + n$. 
\end{enumerate}
\end{fact}

\noindent
We omit the easy proof, because we do not use this result directly.

The factorization constraint~\eqref{eqn:factor-sdr} does not give us direct
access to the factors $\mtx{U}$ and $\mtx{V}$.  Nevertheless, we can place
restrictions on the variables $\mtx{X}$ and $\mtx{Y}$ to limit
the possible values that the factors $\mtx{U}$ and $\mtx{V}$ can take.  If the conditions
are strong enough, it is sometimes possible to determine the factors
completely, modulo symmetries.

\begin{example}[From SVD to eigenvalue decomposition] \label{ex:nuclear}
Let $\mtx{B} \in \R^{n \times m}$ be a matrix.  Consider the semidefinite program
\begin{displaymath}
\underset{\mtx{X} \in \Sym_n, \mtx{Y} \in \Sym_m}{\minimize} \quad \tfrac{1}{2}(\trace(\mtx{X}) + \trace(\mtx{Y}))
\quad\subjto\quad \begin{bmatrix} \mtx{X} & \mtx{B} \\ \mtx{B}^\transp & \mtx{Y} \end{bmatrix} \psdge \mtx{0}.
\end{displaymath}
Every minimizer takes the form $\mtx{X}_\star = \mtx{U} \mtx{\Sigma} \mtx{U}^\transp$
and $\mtx{Y}_{\star} = \mtx{V} \mtx{\Sigma} \mtx{V}^\transp$ where
$\mtx{B} = \mtx{U\Sigma V}^{\transp}$ is a singular value decomposition.
We can find the left and right singular vectors $(\mtx{U}, \mtx{V})$ of $\mtx{B}$
by computing the eigenvalue decompositions of $\mtx{X}_{\star}$ and $\mtx{Y}_{\star}$.
As a side note, the minimal value of the optimization problem
is the Schatten 1-norm (i.e., the sum of singular values) of the matrix $\mtx{B}$.
\end{example}

As we will see, a more elaborate version of the procedure in Example~\ref{ex:nuclear}
allows us to compute an asymmetric sign component decomposition.  To develop
this approach, we require ingredients (Fact~\ref{fact:kt-scd} and Fact~\ref{fact:kt-exposed})
from our work on symmetric sign component decomposition.

\subsection{Overview of algorithm and proof of Theorem~\ref{thm:correct}}
\label{sec:correct-pf}

Given an input matrix $\mtx{B} = \mtx{SW}^\transp$ with a Schur independent sign component $\mtx{S}$,
our aim is to find the (unknown) asymmetric sign component decomposition.
We reduce this challenge to the solved problem of computing a symmetric sign component
decomposition of a correlation matrix.
In this section, we outline the procedure,
along with the proof of Theorem~\ref{thm:correct}.
Algorithm~\ref{alg:asymfactor} encapsulates the computations,
and some details of the argument are postponed to the next sections.

The first step is to construct a correlation matrix
whose symmetric sign component decomposition has the same
sign factor as the input matrix $\mtx{B}$.  To that end,
construct the orthogonal projector $\mtx{P} \in \Sym_n$ onto the
range of $\mtx{B}$.
Then solve the semidefinite program (SDP)
\begin{equation} \label{eqn:factorization-sdp}
\begin{aligned}
&\underset{\mtx{X} \in \Sym_n, \mtx{Y} \in \Sym_m}{\minimize} \quad && \trace(\mtx{Y}) \\
&\subjto\quad && \text{$\trace(\mtx{PX}) = n$ and $\diag(\mtx{X}) = \mathbf{e}$}; \\
&&& \begin{bmatrix} \mtx{X} & \mtx{B} \\ \mtx{B}^\transp & \mtx{Y} \end{bmatrix} \psdge \mtx{0}.
\end{aligned}
\end{equation}
Fact~\ref{fact:factorization} shows that
the semidefinite constraint in~\eqref{eqn:factorization-sdp} links the variables
$\mtx{X}$ and $\mtx{Y}$ to a factorization of $\mtx{B}$.
Meanwhile, courtesy of Fact~\ref{fact:kt-exposed},
the equality constraints in~\eqref{eqn:factorization-sdp}
force the variable $\mtx{X}$ to be a correlation
matrix whose range equals the range of $\mtx{B}$.
The following lemma packages these claims.

\begin{proposition}[Factorization SDP] \label{prop:factorization-sdp}
Instate the assumptions of Theorem~\ref{thm:correct}.  Let $(\mtx{X}_{\star}, \mtx{Y}_{\star})$
be the unique minimizer of the optimization problem~\eqref{eqn:factorization-sdp}.
Then $\mtx{X}_{\star} = \mtx{S}\, \diag(\vct{\tau})\, \mtx{S}^\transp$ where
$\mtx{S} \in \{ \pm 1 \}^{n \times r}$ is the sign component of $\mtx{B}$ and
$\vct{\tau} \in \Delta_r^+$.
\end{proposition}

\noindent
We prove a more detailed version of Proposition~\ref{prop:factorization-sdp}
below in Section~\ref{sec:factorization-sdp}.

The next step is to extract the sign component of the correlation matrix
$\mtx{X}_{\star}$ that solves~\eqref{eqn:factorization-sdp}.
According to Proposition~\ref{prop:factorization-sdp}, the correlation matrix $\mtx{X}_{\star}$
meets the requirements of Fact~\ref{fact:kt-scd}.
Therefore, we can invoke Algorithm~\ref{alg:symfactor},
the symmetric sign component decomposition method,
to obtain a factorization
\begin{displaymath}
\mtx{X}_{\star} = \tilde{\mtx{S}} \, \diag(\tilde{\vct{\tau}}) \, \tilde{\mtx{S}}^\transp
\quad\text{where}\quad
\text{$\tilde{\mtx{S}} = \mtx{S\Pi}$ for a signed permutation $\mtx{\Pi}$.}
\end{displaymath}
We cannot resolve the signed permutation,
but the computed sign component $\tilde{\mtx{S}}$
is equivalent with the designated sign component $\mtx{S}$.

To complete the sign component decomposition,
it remains to determine the weight matrix.
We may do so by solving
the linear system
\begin{displaymath}
\text{find $\tilde{\mtx{W}} \in \R^{m \times r}$}
\quad\subjto\quad
\mtx{B} = \tilde{\mtx{S}} \tilde{\mtx{W}}^\transp.
\end{displaymath}
The solution exists because
$\mtx{B} = \mtx{SW}^\transp = \tilde{\mtx{S}} (\mtx{W} \mtx{\Pi})^\transp$.
The solution $\tilde{\mtx{W}} = \mtx{W} \mtx{\Pi}$ is unique
because $\tilde{\mtx{S}}$ has Schur independent columns,
and so its columns are also linearly independent (Fact~\ref{fact:kt-schur}\eqref{it:schur->linear}).

The pair $(\tilde{\mtx{S}}, \tilde{\mtx{W}})$ yields a sign component
decomposition of the matrix $\mtx{B}$ that is equivalent with the
specified decomposition $\mtx{B} = \mtx{SW}^\transp$.  This
observation completes the proof of Theorem~\ref{thm:correct}.

\subsection{Positive-semidefinite matrices}

It remains to establish Proposition~\ref{prop:factorization-sdp}.
The argument depends on core properties of
psd matrices, which we collect here.  For references,
see~\cite{Bha97:Matrix-Analysis,Bha07:Positive-Definite}.

\begin{fact}[Conjugation rule] \label{fact:psd_invariance}
Conjugation respects the semidefinite order in the following sense.
\begin{enumerate}
\item	If $\mtx{X} \psdge \mtx{0}$, then $\mtx{KXK}^\transp \psdge \mtx{0}$ for each matrix $\mtx{K}$ with compatible dimensions.

\item	If $\mtx{K}$ has full column rank and $\mtx{KXK}^\transp \psdge \mtx{0}$, then $\mtx{X} \psdge \mtx{0}$.
\end{enumerate}
\end{fact}

\begin{fact}[Schur complements] \label{fact:schur_complement}
Assume that $\mtx{X} \in \Sym_n$ is a (strictly) positive-definite matrix.
Then
\begin{displaymath}
\begin{bmatrix} \mtx{X} & \mtx{K} \\ \mtx{K}^\transp & \mtx{Y} \end{bmatrix} \psdge \mtx{0}
\quad\text{if and only if}\quad
\mtx{Y} \psdge \mtx{K}^\transp \mtx{X}^{-1} \mtx{K}.
\end{displaymath}
Related results hold when $\mtx{X}$ is merely psd.
\end{fact}

\begin{fact}[Trace is monotone] \label{fact:trace-monotone}
Let $\mtx{X}$ and $\mtx{Y}$ be psd matrices that satisfy $\mtx{X} \psdge \mtx{Y}$.
Then $\trace(\mtx{X}) \geq \trace(\mtx{Y})$, and equality holds precisely when $\mtx{X} = \mtx{Y}$.
\end{fact}

\subsection{The Factorization SDP}
\label{sec:factorization-sdp}

We are now prepared to prove Proposition~\ref{prop:factorization-sdp},
which describes the solution of the factorization SDP~\eqref{eqn:factorization-sdp}.
The proposition follows instantly from a more precise lemma.

\begin{lemma}[Factorization SDP]
Instate the assumptions of Theorem~\ref{thm:correct}.
Construct the orthogonal projector $\mtx{P} \in \Sym_n$
onto the range of $\mtx{B}$.
Define the positive-definite diagonal matrix
\begin{displaymath}
\mtx{D} = \diag( \norm{ \vct{w}_1 }_{\ell_2}, \dots, \norm{ \vct{w}_r }_{\ell_2} ) \psdgt \mtx{0}
\quad\text{where}\quad
\mtx{W} = \begin{bmatrix} \vct{w}_1 & \dots & \vct{w}_r \end{bmatrix}.
\end{displaymath}
Then the unique solution to the semidefinite optimization problem~\eqref{eqn:factorization-sdp}
is the pair
\begin{displaymath}
\mtx{X}_{\star} = (\trace \mtx{D})^{-1} \, \mtx{SDS}^\transp
\quad\text{and}\quad
\mtx{Y}_{\star} = (\trace \mtx{D}) \, \mtx{WD}^{-1}\mtx{W}^\transp.
\end{displaymath}
\end{lemma}

\begin{proof}
Recall that $\mtx{B} = \mtx{SW}^\transp$ for a Schur independent sign matrix
$\mtx{S}$ and a matrix $\mtx{W}$ with full column rank.

First, we argue that a feasible point $\mtx{X}$
of the factorization SDP~\eqref{eqn:factorization-sdp}
must be a correlation matrix of the form
\begin{equation} \label{eqn:X-feasible}
\mtx{X} = \mtx{S} \, \diag(\vct{\tau}) \, \mtx{S}^\transp
\quad\text{for $\vct{\tau} \in \Delta_r$.}
\end{equation}
Indeed, the block matrix constraint in~\eqref{eqn:factorization-sdp} ensures that $\mtx{X} \psdge \mtx{0}$,
and the constraint $\diag(\mtx{X}) = \mathbf{e}$ makes $\mtx{X}$ a correlation matrix.
At the same time, since the matrix $\mtx{W} \in \R^{m \times r}$ has full column rank,
\begin{displaymath}
\range(\mtx{P}) = \range(\mtx{B}) = \range(\mtx{SW}^\transp) = \range(\mtx{S}).
\end{displaymath}
Fact~\ref{fact:kt-exposed} shows that the constraint $\trace(\mtx{PX}) = n$
isolates the family $\{ \mtx{S}\,\diag(\vct{\tau})\,\mtx{S}^\transp : \vct{\tau} \in \Delta_r \}$
of correlation matrices.  This establishes the claim.

Next, substitute the expression~\eqref{eqn:X-feasible} into the block
matrix constraint in~\eqref{eqn:factorization-sdp} and use the condition $\mtx{B} = \mtx{SW}^\transp$
to factorize:
\begin{displaymath}
\begin{bmatrix} \mtx{X} & \mtx{B} \\ \mtx{B}^\transp & \mtx{Y} \end{bmatrix}
	= \begin{bmatrix} \mtx{S} & \mtx{0} \\ \mtx{0} & \Id \end{bmatrix}
	\begin{bmatrix} \diag(\vct{\tau}) & \mtx{W}^\transp \\ \mtx{W} & \mtx{Y} \end{bmatrix}
	\begin{bmatrix} \mtx{S} & \mtx{0} \\ \mtx{0} & \Id \end{bmatrix}^\transp
	\psdge \mtx{0}.
\end{displaymath}
Since $\mtx{S}$ is Schur independent, it has full column rank (Fact~\ref{fact:kt-schur}\eqref{it:schur->linear}).
Therefore, the conjugation rule (Fact~\ref{fact:psd_invariance}) implies that the psd
constraint in the last display is equivalent with the condition
\begin{equation} \label{eqn:tau-psd}
\begin{bmatrix} \diag(\vct{\tau}) & \mtx{W}^\transp \\ \mtx{W} & \mtx{Y} \end{bmatrix} \psdge \mtx{0}.
\end{equation}
Now, we can recognize that $\diag(\vct{\tau})$ is a strictly positive-definite matrix.
Indeed, owing to~\eqref{eqn:tau-psd}, the relation $\tau_i = 0$ would imply that
the corresponding column $\vct{w}_i$ of the weight matrix equals zero,
but this is impossible because $\mtx{W}$ has full column rank.

Apply the Schur complement rule (Fact~\ref{fact:schur_complement}) to
the matrix~\eqref{eqn:tau-psd} to confirm that
\begin{displaymath}
\mtx{Y} \psdge \mtx{W} \, \diag(\vct{\tau})^{-1} \, \mtx{W}^\transp
	\psdgt \mtx{0}.
\end{displaymath}
The objective function, $\trace(\mtx{Y})$, of the semidefinite program~\eqref{eqn:factorization-sdp}
is strictly monotone with respect to the semidefinite order (Fact~\ref{fact:trace-monotone}).
The variable $\mtx{Y}$ is otherwise unconstrained, so the SDP achieves its minimum if and only if
\begin{displaymath}
\mtx{Y} = \mtx{W} \, \diag(\vct{\tau})^{-1} \, \mtx{W}^\transp.
\end{displaymath}
It remains to determine the vector $\vct{\tau}_{\star} \in \Delta_r^+$
that minimizes the trace of $\mtx{Y}$.

To that end, calculate that
\begin{displaymath}
\trace( \mtx{Y} ) = \trace( \mtx{W} \, \diag(\vct{\tau})^{-1} \, \mtx{W}^\transp )
	= \sum_{i=1}^r \tau_i^{-1} \norm{\vct{w}_i}_{\ell_2}^2
	\geq \min_{1 \leq k \leq r} \big( \tau_k^{-1} \norm{\vct{w}_k}_{\ell_2} \big)
	\cdot \sum_{i=1}^r \norm{\vct{w}_i}_{\ell_2}
	> 0.
\end{displaymath}
Equality holds if and only if the quantities $\tau_k^{-1} \norm{\vct{w}_k}_{\ell_2}$
are identical for all indices $k$.  Since $\vct{\tau} \in \Delta_r^+$, we may
conclude that the minimizer $\vct{\tau}_{\star}$ has coordinates
\begin{displaymath}
(\vct{\tau}_{\star})_k = \norm{\vct{w}_k}_{\ell_2} \left( \sum_{i=1}^r \norm{\vct{w}_i}_{\ell_2} \right)^{-1}
\quad\text{for each index $k$.}
\end{displaymath}

In summary, we have shown that the unique matrices that optimize~\eqref{eqn:factorization-sdp}
take the form
\begin{displaymath}
\mtx{X}_{\star} = \mtx{S} \, \diag(\vct{\tau}_{\star}) \, \mtx{S}^\transp
\quad\text{and}\quad
\mtx{Y}_{\star} = \mtx{W} \, \diag(\vct{\tau}_{\star})^{-1} \, \mtx{W}^\transp.
\end{displaymath}
Identify the diagonal matrix $\mtx{D}$ from the statement to complete the proof. 
\end{proof}

\section{Asymmetric binary component decomposition}
\label{sec:bcd-full}

In this section, we develop a procedure (Algorithm~\ref{alg:binfactor})
for computing an asymmetric binary component
decomposition~\eqref{eqn:bcd-matrix}--\eqref{eqn:bcd-vector}.
We prove Theorem~\ref{thm:bcd-main}, which states that the
algorithm succeeds under a Schur independence condition. 
Our approach reduces the problem of computing a binary component
decomposition to the problem of computing a sign component decomposition
of a related matrix.

\subsection{Correspondence between binary vectors and sign vectors}

As we have discussed, there is a one-to-one correspondence $\mtx{F}$ between
sign vectors and binary vectors~\eqref{eqn:affine-map}.
The correspondence between asymmetric sign component decompositions and binary component decompositions,
however, is more subtle because they are invariant under different transformation.
Indeed, $\vct{sw}^\transp$ does not change if we flip the sign of both $\vct{s} \in \{ \pm 1 \}^n$
and $\vct{w} \in \R^{m}$.  On the other hand, the matrix $\vct{zw}^\transp$ completely
determines the vectors $\vct{z} \in \{0, 1\}^n$ and $\vct{w} \in \R^m$.

\subsection{Reducing binary component decomposition to sign component decomposition}

Given a matrix that has a binary component decomposition, we can apply a simple
transformation to construct a related matrix that admits a sign component decomposition

\begin{proposition}[Binary component decomposition: Reduction] \label{prop:bcd-redux}
Consider a matrix $\mtx{C} \in \R^{n \times m}$ that has a binary component decomposition
\begin{displaymath}
\mtx{C} = \mtx{ZW}^\transp
\quad\text{where}\quad
\text{$\mtx{Z} \in \{0,1\}^{n \times r}$ and $\mtx{W} \in \R^{m \times r}$.}
\end{displaymath}
Construct the matrix 
\begin{displaymath}
\mtx{B} = \mtx{F}(\mtx{C}) = 2\mtx{C} - \mathbf{E} \in \R^{n \times m}.
\end{displaymath}
Then $\mtx{B}$ admits a sign component decomposition with inner dimension $r + 1$:
\begin{equation} \label{eqn:bcd-scd}
\mtx{B} = \begin{bmatrix} \mtx{S} & \mathbf{e} \end{bmatrix}
 \begin{bmatrix} \mtx{W} & \mtx{W}\mathbf{e} - \mathbf{e} \end{bmatrix}^\transp
 \quad\text{where}\quad
 \mtx{S} = \mtx{F}(\mtx{Z}).
\end{equation}
Recall that $\mathbf{E} = \mathbf{ee}^\transp$ is a matrix of ones
with appropriate dimensions.
\end{proposition}

\begin{proof}
The result follows from a straightforward calculation:
\begin{displaymath}
\begin{aligned}
\mtx{B} &= 2\mtx{C} - \mathbf{E} = 2 \mtx{Z} \mtx{W}^\transp - \mathbf{E}
	= (2 \mtx{Z} - \mathbf{E}) \mtx{W}^\transp + \mathbf{E} \mtx{W}^{\transp} - \mathbf{E} \\
	&= \mtx{F}(\mtx{Z}) \mtx{W}^\transp + \mathbf{e} (\mtx{W} \mathbf{e} - \mathbf{e})^\transp
	= \begin{bmatrix} \mtx{F}(\mtx{Z}) & \mathbf{e} \end{bmatrix}
	\begin{bmatrix} \mtx{W} & \mtx{W} \mathbf{e} - \mathbf{e} \end{bmatrix}^\transp.
\end{aligned}
\end{displaymath}
Recognize the matrix $\mtx{S} = \mtx{F}(\mtx{Z})$ to complete the argument.
\end{proof}

\subsection{Resolving the sign ambiguity}

Proposition~\ref{prop:bcd-redux} allows us to reduce the problem of computing
a binary component decomposition to the problem of computing a sign component
decomposition.  Nevertheless, the sign component decomposition has a sign
invariance that is not present in the binary component decomposition.
The next result explains how to resolve this ambiguity.

\begin{proposition}[Sign ambiguity] \label{prop:bcd-sign}
Instate the notation of Proposition~\ref{prop:bcd-redux}.
Assume that the unique sign component decomposition of
$\mtx{B}$ with inner dimension $r+1$ is the one given
in~\eqref{eqn:bcd-scd}, which we write as
\begin{equation} \label{eqn:bcd-scd-factors}
\mtx{B} = \begin{bmatrix} \vct{s}_1 & \dots & \vct{s}_r & \mathbf{e} \end{bmatrix}
	\begin{bmatrix} \vct{w}_1 & \dots & \vct{w}_r & \mtx{W} \mathbf{e} - \mathbf{e} \end{bmatrix}^\transp.
\end{equation}
Suppose that we have computed another sign component decomposition
\begin{displaymath}
\mtx{B} = \tilde{\mtx{S}} \tilde{\mtx{W}}^\transp
	= \begin{bmatrix} \tilde{\vct{s}}_1 & \dots & \tilde{\vct{s}}_r & \mathbf{e} \end{bmatrix}
	\begin{bmatrix} \tilde{\vct{w}}_1 & \dots & \tilde{\vct{w}}_r & \tilde{\vct{w}}_{r+1} \end{bmatrix}^\transp.
\end{displaymath}
By uniqueness, 
there is a signed permutation $(\pi, \vct{\xi})$ on $r$ letters for which
\begin{displaymath}
\tilde{\vct{s}}_i = \xi_i \vct{s}_{\pi(i)}
\quad\text{and}\quad
\tilde{\vct{w}}_i = \xi_i \vct{w}_{\pi(i)}
\quad\text{for each index $i = 1, \dots, r$.}
\end{displaymath}
Then the sign vector $\vct{\xi} \in \R^r$ is the unique solution to the linear system
\begin{displaymath}
\begin{bmatrix} \tilde{\vct{w}}_1 & \dots & \tilde{\vct{w}}_r \end{bmatrix} \vct{\xi}
	= \tilde{\vct{w}}_{r+1} + \mathbf{e}
\end{displaymath}
Moreover, the binary matrix satisfies
\begin{displaymath}
\mtx{Z} = \begin{bmatrix} \vct{z}_1 & \dots & \vct{z}_r \end{bmatrix}
\quad\text{where}\quad
\vct{z}_{\pi(i)} = \mtx{F}^{-1}( \xi_i \tilde{\vct{s}}_i ).
\end{displaymath}
\end{proposition}

\begin{proof}
If the sign component decomposition~\eqref{eqn:bcd-scd-factors} with $r + 1$ terms
is determined up to signed permutation, then $\pm \mathbf{e}$ must appear
among the columns of $\tilde{\mtx{S}}$.  By sign change and permutation,
we may arrange that the last column of $\tilde{\mtx{S}}$ equals $\mathbf{e}$.
The simultaneous transformation on $\tilde{\mtx{W}}$ ensures that its
last column satisfies $\tilde{\vct{w}}_{r+1} = \mtx{W} \mathbf{e} - \mathbf{e}$.
The uniqueness of the decomposition~\eqref{eqn:bcd-scd-factors} ensures that
the weight matrix has full column rank, and so $\tilde{\mtx{W}}$ also has full column rank.
Now, observe that
\begin{displaymath}
\begin{bmatrix} \tilde{\vct{w}}_1 & \dots & \tilde{\vct{w}}_r \end{bmatrix} \vct{\xi}
	= \begin{bmatrix} \xi_1 \vct{w}_{\pi(1)} & \dots & \xi_r \vct{w}_{\pi(r)} \end{bmatrix} \vct{\xi}
	= \sum_{i=1}^r \xi_i^2 \vct{w}_{\pi(i)}
	= \sum_{i=1}^r \vct{w}_i
	= \mtx{W} \mathbf{e}
	= \tilde{\vct{w}}_{r+1} + \mathbf{e}.
\end{displaymath}
The solution to this linear system is uniquely determined,
so it must equal the true vector $\vct{\xi}$ of sign flips.  Therefore,
\begin{displaymath}
\vct{z}_{\pi(i)} = \mtx{F}^{-1}(\vct{s}_{\pi(i)}) = \mtx{F}^{-1}( \xi_i \tilde{\vct{s}}_i )
\quad\text{for each index $i = 1, \dots, r$.}
\end{displaymath}
This completes the argument.
\end{proof}

\subsection{Uniqueness}

Next, we confirm that binary component decompositions are unique
under a Schur independence condition. 
The key step is to argue that ordinary permutations are the only invertible transformations
that preserve Schur independence of binary vectors.
This result is an analog of Proposition~\ref{prop:schur-transform},
and the proof follows a similar pattern.

\begin{proposition}[Binary Schur independence: Transformations] \label{prop:bin-schur-transform}
Let $\mtx{Z} \in \{ 0, 1 \}^{n \times r}$ be a Schur independent binary matrix,
and let $\mtx{Q} \in \R^{r \times r}$ be an invertible matrix.
Then $\mtx{ZQ}$ is a binary matrix if and only if $\mtx{Q}$ is a permutation matrix.
\end{proposition}

\begin{proof}
If $\mtx{Q}$ is a permutation, then it is clear that $\mtx{ZQ}$ is a binary matrix.
Let us prove the converse.

Write $\vct{z}_i$ for the columns of $\mtx{Z}$;
write $\vct{q}_i$ for the columns of $\mtx{Q}$;
and write $\tilde{\vct{z}}_i$ for the columns of $\mtx{ZQ}$.
For each index $1 \leq k \leq r$,
\begin{displaymath}
\tilde{\vct{z}}_k = \mtx{Z} \vct{q}_k = \sum_{i=1}^r \ip{ \mathbf{e}_i }{ \vct{q}_k }\, \vct{z}_i.
\end{displaymath}
The vector $\tilde{\vct{z}}_k$ is binary, so
\begin{displaymath}
\begin{aligned}
\vct{0} = (\mathbf{e} - \tilde{\vct{z}}_k) \odot \tilde{\vct{z}}_k
	&= \sum_{i=1}^r \ip{ \mathbf{e}_i }{ \vct{q}_k } (\mathbf{e} \odot \vct{z}_i)
	- \sum_{i,j=1}^r \ip{ \mathbf{e}_i }{ \vct{q}_k } \ip{ \mathbf{e}_j }{ \vct{q}_k }( \vct{z}_i \odot \vct{z}_j ) \\ 
	&= \sum_{i=1}^r \ip{ \mathbf{e}_i }{ \vct{q}_k } (1 - \ip{ \mathbf{e}_i }{ \vct{q}_k }) \, \vct{z}_i
	-2 \sum_{i < j} \ip{ \mathbf{e}_i }{ \vct{q}_k } \ip{ \mathbf{e}_j }{ \vct{q}_k }( \vct{z}_i \odot \vct{z}_j ).
\end{aligned}
\end{displaymath}
We have used the fact that $\vct{z}_i = \mathbf{e} \odot \vct{z}_i = \vct{z}_i \odot \vct{z}_i$ for each binary vector.
Schur independence of the matrix $\mtx{Z}$ ensures that the vectors on the right-hand side
of this expression compose a linearly independent family.  It follows that
\begin{displaymath}
\ip{ \mathbf{e}_i }{ \vct{q}_k } ( 1 - \ip{ \mathbf{e}_i }{ \vct{q}_k }) = 0
\quad\text{and}\quad
\ip{ \mathbf{e}_i }{ \vct{q}_k } \ip{ \mathbf{e}_j }{ \vct{q}_k } = 0
\quad\text{when $i \neq j$.}
\end{displaymath}
Therefore, $\vct{q}_k$ must be a standard basis vector:
$\vct{q}_k = \mathbf{e}_{\pi(k)} \in \R^r$ for an index $\pi(k) \in \{ 1, \dots, r \}$.
Since the matrix $\mtx{Q}$ is invertible, $\pi$ is a permutation on $r$ letters.
In other words, $\mtx{Q}$ is a permutation matrix.
\end{proof}

As a consequence, we obtain a result about the uniqueness of
asymmetric binary component decompositions, modulo permutation.

\begin{theorem}[Binary component decomposition: Uniqueness] \label{thm:bin-uniqueness}
Consider a matrix $\mtx{C} \in \R^{n \times m}$
that admits a binary component decomposition $\mtx{C} = \mtx{ZW}^\transp$.
Assume that
\begin{enumerate}
\item	The binary matrix $\mtx{Z} \in \{ 0, 1\}^{n \times r}$ is Schur independent;
\item	The weight matrix $\mtx{W} \in \R^{m \times r}$ has full column rank.
\end{enumerate}
Then all minimal sign component decompositions of $\mtx{C}$ (with inner dimension $r$)
are equivalent, up to simultaneous permutation of the columns of the factors.
\end{theorem}

\noindent
We omit the proof, which mirrors the argument in Theorem~\ref{thm:bin-uniqueness}.

\subsection{Computation}
\label{sec:bcd-compute}

Proposition~\ref{prop:bcd-redux} and Proposition~\ref{prop:bcd-sign} give us a mechanism
for computing a binary component decomposition, provided that an associated matrix has
a unique sign component decomposition.
We can exploit our theory on the tractable computation of sign component decompositions
to identify situations where we can compute binary component decompositions.

\begin{theorem}[Binary component decomposition: Computation] \label{thm:bcd-correct}
Consider a matrix $\mtx{C} \in \R^{n \times m}$ that admits
a sign component decomposition $\mtx{C} = \mtx{ZW}^\transp$.
Assume that
\begin{enumerate}
\item	The sign matrix $\mtx{Z} \in \{0,1 \}^{n \times r}$ is Schur independent;
\item	The weight matrix $\mtx{W} \in \R^{m \times r}$ has full column rank.
\end{enumerate}
Then, with probability one, Algorithm~\ref{alg:binfactor} identifies
the minimal binary component decomposition. 
\end{theorem}

\begin{proof}[Proof of Theorem~\ref{thm:bcd-correct}]
Suppose that $\mtx{C} \in \R^{n \times m}$ has a binary component decomposition
$\mtx{C} = \sum_{i=1}^r \vct{z}_i \vct{w}_i^\transp$ involving a Schur independent
family $\{ \vct{z}_1, \dots, \vct{z}_r \} \subseteq \{ 0, 1\}^r$ of binary components.
Introduce the associated sign vectors $\vct{s}_i = \mtx{F}(\vct{z}_i)$.
By Fact~\ref{fact:bcd-scd}, the family $\{ \vct{s}_1, \dots, \vct{s}_r, \mathbf{e} \}$
of sign vectors is Schur independent.

Proposition~\ref{prop:bcd-redux} shows that the matrix
$\mtx{B} = 2\mtx{C} - \mathbf{E} = \begin{bmatrix} \mtx{S} & \mathbf{e} \end{bmatrix}
\begin{bmatrix} \mtx{W} & \mtx{W}\mathbf{e} - \mathbf{e} \end{bmatrix}^\transp$.
Define $\vct{s}_{r+1} = \mathbf{e}$ and $\vct{w}_{r+1} = \mtx{W}\mathbf{e} - \mathbf{e}$.
By Theorem~\ref{thm:correct}, Algorithm~\ref{alg:asymfactor}
allows us to compute pairs $(\tilde{\vct{s}}_i, \tilde{\vct{w}}_i)$
with the property that $\tilde{\vct{s}}_i = \xi_i \vct{s}_{\pi(i)}$
and $\tilde{\vct{w}}_i = \xi_i \vct{w}_{\pi(i)}$ for each $i = 1, \dots, r+1$
where $(\pi, \vct{\xi})$ is a signed permutation.
By change of sign and permutation,
we can assume that $\tilde{\vct{s}}_{r+1} = \mathbf{e}$ and
$\tilde{\vct{w}}_{r+1} = \mtx{W}\mathbf{e} - \mathbf{e}$.
Proposition~\ref{prop:bcd-sign} shows that we can use
these computed weight vectors $\tilde{\vct{w}}_i$ to find the
sign vector $\vct{\xi}$, and we obtain the binary components of $\mtx{C}$ as
$\vct{z}_{\pi(i)} = \tilde{\vct{z}}_i = \mtx{F}^{-1}(\xi_i \tilde{\vct{s}}_i)$.

Last, we define the binary factor matrix
$\tilde{\mtx{Z}} = \begin{bmatrix} \tilde{\vct{z}}_1 & \dots & \tilde{\vct{z}}_r \end{bmatrix}$
and construct a weight matrix with the correct signs:
$\tilde{\mtx{W}}_{+} = \begin{bmatrix} \xi_1 \tilde{\vct{w}}_1 & \dots & \xi_r \tilde{\vct{w}}_r \end{bmatrix}$.
\end{proof}

\section{Robustness}
\label{sec:robustness}

In this paper, we have designed factorization algorithms
to operate on low-rank matrices that admit an exact
sign component decomposition or binary component decomposition.
In most practical applications, however, the data matrix
will be contaminated by noise, errors, or outliers.
One way to handle these non-idealities is to process
the data to remove corruptions.  Afterward,
we can apply our algorithms to factorize the clean data matrix.
This section describes some situations where we can
implement this idea.

\subsection{Approximate factorization of a noisy matrix}

We focus on computing an asymmetric sign component decomposition
of a noisy data matrix:
\begin{displaymath}
\mtx{B} = \mtx{L}_0 + \mtx{\Omega}_0 \in \R^{n \times m}
\quad\text{where}\quad
\mtx{L}_0 = \mtx{SW}^\transp.
\end{displaymath}
As usual, the factor $\mtx{S} \in \{ \pm 1 \}^{n \times r}$
and $\mtx{W} \in \R^{m \times r}$.  The matrix
$\mtx{\Omega}_0 \in \R^{n \times m}$ captures the noise.

Given the matrix $\mtx{B}$,
our computational goal is to remove the noise $\mtx{\Omega}_0$ completely.
Then we can factorize the clean matrix $\mtx{L}_0$ that remains.
In summary,

\begin{enumerate}
\item	\textbf{Denoise:}  Remove the noise $\mtx{\Omega}_0$ from the observed matrix $\mtx{B}$
to obtain the clean data $\mtx{L}_0$.

\item	\textbf{Factorize:}  Compute a sign component decomposition
$\mtx{L}_0 = \tilde{\mtx{S}} \tilde{\mtx{W}}^\transp$.
\end{enumerate}
We can realize this strategy for several different noise models.  In particular,

\begin{itemize}
\item	\textbf{Gross errors in entries:}  The corruption $\mtx{\Omega}_0$ is a sparse matrix.
This model is appropriate for errors in individual measurements.

\item	\textbf{Gross errors in columns:}  The corruption $\mtx{\Omega}_0$ is a column-sparse matrix.
This model is appropriate for handling outliers.
\end{itemize}

\noindent
To facilitate the analysis, we also instate simple probabilistic models
for the clean data matrix $\mtx{L}_0$ and for the noise $\mtx{\Omega}_0$.
These assumptions can be relaxed substantially.

\subsection{The Gaussian loadings model}

We work with a generative probabilistic model for the clean data matrix.
This model combines a fixed set of sign vectors with random coefficients.

\begin{model}[Gaussian loadings] \label{mod:glm}
Select a Schur-independent sign matrix $\mtx{S} \in \{ \pm 1 \}^{n \times r}$
and a natural number $m \geq r$.  The Gaussian loading model
$\mathrm{GLM}(\mtx{S}, m)$ is a distribution on matrices in $\R^{n \times m}$.
A matrix $\mtx{L}_0$ drawn from this model takes the form
\begin{displaymath}
\mtx{L}_0 = \frac{1}{\sqrt{r}} \mtx{SG}^\transp
\quad\text{where}\quad
\text{$\mtx{G} \in \R^{m \times r}$ is standard normal.}
\end{displaymath}
\end{model}

Owing to Theorem~\ref{thm:scd-main}, a realization $\mtx{L}_0 \sim \mathrm{GLM}(\mtx{S}, m)$
almost surely has a unique sign component decomposition.  Moreover, we can produce
an equivalent decomposition $\mtx{L}_0 = \tilde{\mtx{S}} \tilde{\mtx{W}}^\transp$
using Algorithm~\ref{alg:asymfactor}.

Suppose we are given a noisy observation of a matrix drawn from the Gaussian loadings model: 
\begin{displaymath}
\mtx{B} = \mtx{L}_0 + \mtx{\Omega}_0
\quad\text{where}\quad
\mtx{L}_0 \sim \mathrm{GLM}(\mtx{S}, m).
\end{displaymath}
We are interested in removing the noise.
The difficulty of this problem depends on the choice of the sign matrix $\mtx{S}$,
the number $m$ of independent data points that we sample, and the type
of noise $\mtx{\Omega}_0$ that we must contend with.

Let us explain how the geometry of the Gaussian loadings model enters the picture.
The sign matrix $\mtx{S}$ determines the anisotropy of the columns of $\mtx{L}_0$.
Indeed,
\begin{displaymath}
\Expect[ \mtx{L}_0 \mtx{L}_0^\transp ] = \frac{1}{r} \Expect[ \mtx{S} \mtx{GG}^\transp \mtx{S}^\transp ]
	= \frac{m}{r} \cdot \mtx{SS}^\transp \in \Sym_n.
\end{displaymath}
We can summarize how well the columns of $\mtx{L}_0$ fill out their
span by means of the \emph{permeance statistic}:
\begin{displaymath}
\nu(\mtx{S}) = \frac{1}{n} \lambda_{\min}( \mtx{S}^\transp \mtx{S} )
	= \frac{1}{n} \min_{\norm{\vct{u}}_{\ell_2} = 1} \norm{\mtx{S}\vct{u}}_{\ell_2}^2
	\in [0, 1].
\end{displaymath}
If the columns of $\mtx{S}$ are orthogonal, then $\nu(\mtx{S}) = 1$;
if the columns of $\mtx{S}$ are strongly aligned, then $\nu(\mtx{S}) \approx 0$.
Given noisy data, finding all of the directions in the range of $\mtx{S}$
is harder when the permeance is small.

\subsection{Gross errors in matrix entries}

In this section, we consider a noise model in which a moderate number of entries
of the data matrix are corrupted arbitrarily.  Suppose that we observe
\begin{equation} \label{eqn:rank-sparsity}
\mtx{B} = \mtx{L}_0 + \mtx{\Omega}_0 \in \R^{n \times m},
\end{equation}
where $\mtx{L}_0$ is a low-rank data matrix, while $\mtx{\Omega}_0$ models
contamination of individual entries.  We imagine that $\mtx{\Omega}_0$ is sparse,
but its nonzero entries are arbitrary.  In particular, the noise can erase
individual entries of the data matrix or change them maliciously.

The data model~\eqref{eqn:rank-sparsity} was first studied
by Chandrasekaran et al.~\cite{CSPW11:Rank-Sparsity}.
Given the observation $\mtx{B}$, they proposed to separate
the low-rank data matrix $\mtx{L}_0$ from the noise $\mtx{\Omega}_0$
by solving a semidefinite optimization problem:
\begin{equation} \label{eqn:rank-sparsity-opt}
\begin{aligned}
&\underset{\mtx{L}, \mtx{\Omega} \in \R^{n \times m}}{\minimize} &&
	\norm{\mtx{L}}_{S_1} + \frac{1}{\sqrt{\max\{n,m\}}} \norm{\mtx{\Omega}}_{\ell_1}
&\subjto && \mtx{B} = \mtx{L} + \mtx{\Omega}.
\end{aligned}
\end{equation}
The Schatten 1-norm $\norm{\cdot}_{S_1}$ promotes low-rank in the component $\mtx{L}$,
while the vectorized $\ell_1$ norm $\norm{\cdot}_{\ell_1}$ promotes sparsity
in the other component $\mtx{\Omega}$.

Chandrasekaran et al.~\cite{CSPW11:Rank-Sparsity} developed deterministic
conditions under which the pair $(\mtx{L}_0, \mtx{\Omega}_0)$ is the unique
solution of the optimization problem~\eqref{eqn:rank-sparsity-opt}.
The subsequent paper~\cite{CLMW11:RobustPCA} established additional guarantees
under a probabilistic model for the data.  By adapting these arguments,
we can show that it is often possible to remove sparse errors from
a realization of the Gaussian loadings model.  We have the following result.

\begin{theorem}[Gaussian loadings model: Sparse noise] \label{thm:glm-sparse}
Suppose we observe the  matrix
$\mtx{B} = \mtx{L}_0 + \mtx{\Omega}_0 \in \R^{n \times m}$
where
\begin{enumerate}
\item	The sign matrix $\mtx{S} \in \{\pm 1\}^{n \times r}$ is Schur independent, with permeance $\nu(\mtx{S})$;
\item	The data matrix $\mtx{L}_0 \sim \mathrm{GLM}(\mtx{S}, m)$ is drawn from the Gaussian loadings model; and
\item	The noise matrix $\mtx{\Omega}_0$ has $\omega$ nonzero entries, in uniformly random locations,
with arbitrary magnitudes that may depend on $\mtx{L}_0$.
\end{enumerate}
Assume that the parameters satisfy
\begin{equation} \label{eqn:rank-sparsity-cond}
\nu(\mtx{S}) \gtrsim \frac{r \log^3(n+m)}{\min\{n,m\}}
\quad\text{and}\quad
\omega \lesssim mn.
\end{equation}
Then, with high probability over the randomness in the model,
the pair $(\mtx{L}_0, \mtx{\Omega}_0)$ is the unique solution
to the optimization problem~\eqref{eqn:rank-sparsity-opt}.
\end{theorem}

\noindent
See Appendix~\ref{app:rank-sparsity} for the proof, which relies
on standard methods from high-dimensional probability.
Better results may be possible with additional argument.

Let us take a moment to discuss Theorem~\ref{thm:glm-sparse}.
The condition~\eqref{eqn:rank-sparsity-cond} on the sparsity $\omega$
of the noise allows us to repair errors in a constant proportion
of the entries of the data matrix.
The condition~\eqref{eqn:rank-sparsity-cond} on the permeance $\nu(\mtx{S})$
states that the conditioning of the sign matrix $\mtx{S}$ controls
the minimum number $m$ of samples that we need to remove the noise.
It also gives a minimum standard on the conditioning, in terms
of the dimension $n$, for the theorem to operate.

\subsection{Gross errors in matrix columns}

In this section, we study a noise model in which a moderate number
of columns of the data matrix are corrupted arbitrarily.
Suppose that we observe a matrix of the form
\begin{equation} \label{eqn:in-n-out}
\mtx{B} = \begin{bmatrix} \mtx{L}_0 & \mtx{\Omega}_0 \end{bmatrix} \mtx{\Pi} \in \R^{n \times M}
\end{equation}
where $\mtx{L}_0 \in \R^{n \times m}$ is a rank-$r$ matrix of clean data, $\mtx{\Omega}_0 \in \R^{n \times m'}$
contains arbitrary noise, $\mtx{\Pi} \in \R^{M \times M}$ is an unknown permutation,
and $M = m + m'$.  We regard the columns of $\mtx{\Omega}_0$ as outliers that are mixed in with
the data and that we need to remove.  The number of inliers ($m$) and outliers $(m')$
is not known in advance, but we do require the value $r$ of the rank.

The data model~\eqref{eqn:in-n-out} has a long history in robust statistics~\cite{HR09:Robust-Statistics}.
In recent years, researchers have proposed a number of methods for removing outliers by means of convex
optimization.  We outline a technique that is based
on the following intuition~\cite{LMTZ15:Robust-Computation}.
Observe that the orthogonal projector $\mtx{P}_0 \in \Sym_n$
onto the range of $\mtx{L}_0$ discriminates inliers from outliers.  Indeed,
$\mtx{P}_0 \vct{x} = \vct{x}$ for $\vct{x} \in \range(\mtx{L}_0)$,
while $\mtx{P}_0 \vct{x} \neq \vct{x}$ for $\vct{x} \notin \range(\mtx{L}_0)$.
Therefore, we may try to find a rank-$r$ projector that fixes many columns of
$\mtx{B}$ by charging as much as possible for columns that are not reproduced.

If we can determine the projector $\mtx{P}_0$, then we can find the low-rank matrix
$\mtx{L}_0$ by picking out the columns of $\mtx{B}$ that are fixed by the projector $\mtx{P}_0$.
(It is possible that $\mtx{P}_0$ also fixes some columns of the noise matrix $\mtx{\Omega}_0$,
but then we should probably regard these outliers as inliers.)

Lerman et al.~\cite{LMTZ15:Robust-Computation} propose to find
the projector $\mtx{P}_0$ by solving the semidefinite optimization problem
\begin{equation} \label{eqn:reaper}
\begin{aligned}
&\underset{\mtx{P} \in \Sym_n}{\minimize} &&
\sum_{i=1}^{M} \norm{ (\Id - \mtx{P}) \mtx{B} \mathbf{e}_i }_{\ell_2}
&\subjto &&
\text{$\trace(\mtx{P}) = r$ and $\mtx{0} \psdle \mtx{P} \psdle \Id$.}
\end{aligned}
\end{equation}
The constraint set is the best convex relaxation of the set of rank-$r$
orthogonal projectors.

Lerman et al.~\cite{LMTZ15:Robust-Computation} develop deterministic
conditions under which the orthogonal projector $\mtx{P}_0$
onto $\range(\mtx{L}_0)$ is the unique solution
to the problem~\eqref{eqn:reaper},
and they specialize these results to random data models.
By adapting their arguments,
we can establish that it is possible to remove outliers
from a data matrix that follows the Gaussian loadings model.

\begin{theorem}[Gaussian loadings model: Outliers] \label{thm:glm-outliers}
Suppose we observe the matrix
$\mtx{B} = \begin{bmatrix} \mtx{L}_0 & \mtx{\Omega}_0 \end{bmatrix} \mtx{\Pi}
\in \R^{n \times M}$,
where $M = m + m'$ and
\begin{enumerate}
\item	The sign matrix $\mtx{S} \in \{\pm 1\}^{n \times r}$ is Schur independent, with permeance $\nu(\mtx{S})$;
\item	The data matrix $\mtx{L}_0 \sim \mathrm{GLM}(\mtx{S}, m)$ is drawn from the Gaussian loadings model; 
\item	The noise matrix $\mtx{\Omega}_0 \in \R^{n \times m'}$ is standard normal; and
\item	The permutation matrix $\mtx{\Pi} \in \R^{M \times M}$ is arbitrary.
\end{enumerate}
Assume that the parameters satisfy
\begin{equation} \label{eqn:reaper-cond}
\max\left\{ 1, \frac{m'}{n} \right\}
\lesssim \sqrt{ \nu(\mtx{S}) } \cdot \frac{m}{r}.
\end{equation}
Then, with high probability over the randomness in the model,
the orthogonal projector onto $\range(\mtx{L}_0)$ is the unique
solution to the optimization problem~\eqref{eqn:rank-sparsity-opt}.
\end{theorem}

\noindent
The proof of Theorem~\ref{thm:glm-outliers} appears in Appendix~\ref{app:glm-outliers}.
Results for other data models are also possible.

Let us discuss Theorem~\ref{thm:glm-outliers} briefly.
First, under the data model,
the columns of $\mtx{L}_0$ and $\mtx{\Omega}_0$ all have
the same energy, on average, so we cannot distinguish
the inliers and the outliers on the basis of their norm.
Next, the condition in~\eqref{eqn:reaper-cond} requires that the number $m'$ of outliers
relative to the ambient dimension $n$ is no greater than the number $m$ of inliers
relative to the dimension $r$ of the space spanned by the inliers.  The permeance
$\nu(\mtx{S})$ also affects how many outliers we can tolerate; it is easier to
reject outliers when the columns of $\mtx{L}_0$ are well distributed.

\subsection{Future work}

Our results on robustness (Theorem~\ref{thm:glm-sparse} and Theorem~\ref{thm:glm-outliers})
are based on the fact that subspaces have a very strong signal,
so it is easy to reject noise that violates the subspace structure.
We can exploit the same intuition to handle some other types of
noise models.  On the other hand, these ideas do not suffice to
treat robustness of the sign component decomposition with respect
to, say, additive Gaussian noise.
An important direction for future research is to develop
additional tools for producing a sign component decomposition
of a noisy data matrix drawn from a more general model.

\appendix

\section{Gaussian loadings model: Removing gross errors in matrix entries}
\label{app:rank-sparsity}

\subsection{Background} 

We consider a noise model in which a moderate number of entries of a data matrix are corrupted arbitrarily. We observe
\begin{displaymath}
\mtx{B} = \mtx{L}_0 + \mtx{\Omega}_0 \in \mathbb{R}^{n \times m},
\end{displaymath}
where $\mtx{L}_0 \in \mathbb{R}^{n \times m}$ is a low-rank data matrix and $\mtx{\Omega}_0 \in
\mathbb{R}^{n \times m}$ models contamination of individual entries. 
We assume that $\mtx{\Omega}_0$ is sparse and both contributions to $\mtx{B}$ are well separated in the following sense. The low rank data matrix $\mtx{L}_0$ cannot be too sparse, while the sparse noise corruption $\mtx{\Omega}_0$ cannot be too low-rank.
Choosing the support of $\mtx{\Omega}_0$ uniformly at random takes care of the latter condition.
The \emph{incoherence statistics} has been identified as a witness for the former.
Set $r = \mathrm{rank}(\mtx{L}_0)$ and let $\mtx{L}_0=\mtx{U} \mtx{\Sigma} \mtx{V}^\transp$ be a SVD \eqref{eqn:svd-matrix}. Note that the orthonormal matrices $\mtx{U} \in \mathbb{R}^{n \times r}$ and $\mtx{V} \in \mathbb{R}^{m \times r}$ are closely related to the projectors onto row- and column range of $\mtx{L}_0$. More precisely, $\mtx{P}_{\mathrm{ran}(\mtx{L}_0)}=\mtx{U} \mtx{U}^\transp \in \Sym_n$ and $\mtx{P}_{\mathrm{ran}(\mtx{L}_0^\transp)} = \mtx{V} \mtx{V}^\transp \in \Sym_m$.
Based on the SVD, we define three \emph{incoherence parameters}:
\begin{equation} \label{eq:incoherence_parameters}
\begin{aligned}
\mu_{(l)} \left( \mtx{L}_0\right) =& \frac{n}{r}\max_{1 \leq k \leq n} \left\|
\mtx{P}_{\mathrm{ran}(\mtx{L}_0)} \mathbf{e}_k \right\|_{\ell_2}^2 = \frac{n}{r} \max_{1 \leq k \leq n} \left\| \mtx{U}^\transp \mathbf{e}_k \right\|_{\ell_2}^2, \\
\mu_{(r)} \left(\mtx{L}_0 \right) =& 
 \frac{m}{r} \max_{1 \leq l \leq m} \left\| \mtx{P}_{\mathrm{ran}(\mtx{L}_0^\transp)} \mathbf{e}_l
\right\|_{\ell_2}^2 = \frac{m}{r} \max_{1 \leq l \leq m} \left\| \mtx{V}^\transp
\mathbf{e}_l \right\|_{\ell_2}^2, \\
\tilde{\mu}(\mtx{L}_0) =& \frac{nm}{r} \max_{1 \leq k \leq n, 1 \leq l \leq m}
\left| \langle \mathbf{e}_k, \mtx{U} \mtx{V}^\transp \mathbf{e}_l \rangle \right|^2.
\end{aligned}
\end{equation}
The first two parameters have a ready explanation: they measure how well spread-out the
left- and right- singular vectors are with respect to the standard basis.
Small values ensure that these vectors are not (too) sparse.
The third parameter lacks a compelling interpretation. It
should be viewed as a technical requirement that features prominently in the arguments by Cand\`es et al.~\cite{CLMW11:RobustPCA}. 
Define the maximum of these three parameters
\begin{equation} \label{eq:incoherence-appendix}
\mu (\mtx{L}_0) = \max \left\{ \mu_{(l)} (\mtx{L}_0), \mu_{(r)} (\mtx{L}_0),
\tilde{\mu}(\mtx{L}_0)\right\}
\end{equation}
and consider the following semidefinite program for rank-sparsity decomposition \cite{CSPW11:Rank-Sparsity}:
\begin{equation} \label{eq:rank_sparsity_decomposition_appendix}
\begin{aligned}
&\underset{\mtx{L}, \mtx{\Omega} \in \R^{n \times m}}{\minimize} &&
	\norm{\mtx{L}}_{S_1} + \tfrac{1}{\sqrt{\max\{n,m\}}} \norm{\mtx{\Omega}}_{\ell_1}
&\subjto && \mtx{B} = \mtx{L} + \mtx{\Omega}.
\end{aligned}
\end{equation}
The main result \cite[Thm.~1]{CLMW11:RobustPCA} establishes a probabilistic recovery guarantee for rank-sparsity decomposition.

\begin{theorem}[Cand\`es, Li, Ma \& Wright] \label{thm:robust_pca}
Suppose that $\mtx{B}=\mtx{L}_0 + \mtx{\Omega}_0 \in \mathbb{R}^{n \times m}$ is such that
\begin{enumerate}
\item $\mtx{L}_0$ is such that rank $r=\mathrm{rank}(\mtx{L}_0)$ and incoherence $\mu (\mtx{L}_0)$ \eqref{eq:incoherence-appendix} obey 
\begin{displaymath}
r \leq \rho_{(r)} \frac{\min \left\{n,m\right\}}{ \mu (\mtx{L}_0) \log^{2}(n+m)};
\end{displaymath} 
\item $\mtx{\Omega}_0 \in \mathbb{R}^{n \times m}$ has $\omega$ nonzero entries, in uniformly random locations, with arbitrary magnitudes that may depend on $\mtx{L}_0$.
\end{enumerate}
Then, with probability at least $1-\gamma \max \left\{n,m\right\}^{-10}$ (over the
choice of the sparsity pattern), the solution to the semidefinite program \eqref{eq:rank_sparsity_decomposition_appendix} is $(\mtx{L}_\star,\mtx{\Omega}_\star) = \left( \mtx{L}_0, \mtx{\Omega}_0 \right)$.
Here, $\rho_{(r)},\rho_{(s)}$ and $\gamma$ are constants of appropriate
 size.
\end{theorem}

\subsection{Coherence statistics for the Gaussian loadings model and denoising}

Recall the \emph{Gaussian loadings model} (Model~\ref{mod:glm}). Select a Schur-independent sign matrix $\mtx{S} \in \left\{ \pm 1 \right\}^{n \times r}$ and a number $m \geq r$. A matrix $\mtx{L}_0 \sim \mathrm{GLM}(\mtx{S},m)$ takes the form $\mtx{L}_0 = r^{-1/2} \mtx{S} \mtx{G}^\transp$, where $\mtx{G} \in \mathbb{R}^{m \times r}$ is standard normal. 
The \emph{permeance statistic} captures how well the columns of $\mtx{L}_0$ fill out their span:
\begin{equation} \label{eq:permeance-appendix}
\nu(\mtx{S}) = \frac{1}{n} \lambda_{\min}( \mtx{S}^\transp \mtx{S} )
	= \frac{1}{n} \min_{\norm{\vct{u}}_{\ell_2} = 1} \norm{\mtx{S}\vct{u}}_{\ell_2}^2.
\end{equation}
The following technical statement implies that the permeance also controls two of the three incoherence parameters \eqref{eq:incoherence_parameters} associated with the Gaussian loadings model.

\begin{lemma} \label{lem:GLM_coherence}
Fix $\mtx{S} \in \left\{\pm 1 \right\}^{n\times r}$ with permeance $\nu (\mtx{S}) >0$ and a natural number $m \geq r$.
Choose $\mtx{L}_0$ from the Gaussian loadings model. Then, for any $\alpha >0$,
\begin{displaymath}
\begin{aligned}
\mu_{(l)} \left( \mtx{L}_0\right) \leq &\nu (\mtx{S})^{-1}, \\
\mathrm{Pr} \left[ \mu_{(r)} \left( \mtx{L}_0\right) \geq 4(\alpha+1) \log (n+m) \right]
\leq & 2(nm)^{-\alpha}/n, \\
\mathrm{Pr} \left[ \tilde{\mu}(\mtx{L}_0) \geq \tfrac{4 (\alpha+1)}{\nu (\mtx{S})}\log
(n+m)\right] \leq & 2 (nm)^{-\alpha}.
\end{aligned}
\end{displaymath}
\end{lemma}
\noindent
The first bound is deterministic and original. The remaining probabilistic bounds follow from adapting arguments by Cand\`es et al.\ 
\cite[Sec.~2.2]{CLMW11:RobustPCA} to the Gaussian loadings model. 
Inserting these bounds into \cite[Thm.~1]{CLMW11:RobustPCA} readily implies that  gross errors in entries of a Gaussian loadings matrix can be removed completely.

\begin{theorem}[Theorem~\ref{thm:glm-sparse}, restatement] \label{thm:glm-sparse-restatement}
Suppose we observe the  matrix
$\mtx{B} = \mtx{L}_0 + \mtx{\Omega}_0 \in \R^{n \times m}$
where
\begin{enumerate}
\item	The sign matrix $\mtx{S} \in \{\pm 1\}^{n \times r}$ is Schur independent, with permeance $\nu(\mtx{S})$;
\item	The data matrix $\mtx{L}_0 \sim \mathrm{GLM}(\mtx{S}, m)$ is drawn from the Gaussian loadings model; and
\item	The noise matrix $\mtx{\Omega}_0$ has $\omega$ nonzero entries, in uniformly random locations,
with arbitrary magnitudes that may depend on $\mtx{L}_0$.
\end{enumerate}
Assume that the parameters satisfy $\omega \leq mn$ and
\begin{equation} \label{eq:glm-sparse-condition}
\nu(\mtx{S}) \gtrsim \frac{r \log^3(n+m)}{\min\{n,m\}}
\end{equation}
Then, with high probability over the randomness in the model,
the pair $(\mtx{L}_0, \mtx{\Omega}_0)$ is the unique solution
to the optimization problem~\eqref{eq:rank_sparsity_decomposition_appendix}.
\end{theorem}

\begin{proof}
Sample $\mtx{L}_0$ from the Gaussian loadings model $\mathrm{GLM}(\mtx{S},m)$. Choose $\alpha=1$ and combine Lemma~\ref{lem:GLM_coherence} with a union bound to conclude
\begin{displaymath}
\mu (\mtx{L}_0) = \max \left\{ \mu_{(r)} (\mtx{L}_0),\mu_{(l)} (\mtx{L}_0), \tilde{\mu}(\mtx{L}_0) \right\} \leq \tfrac{8}{\nu (\mtx{S})} \log (n+m)
\end{displaymath}
with probability at least $ 1-\tfrac{4}{nm}$.
Conditioned on this event, condition (1) in Theorem~\ref{thm:robust_pca} becomes equivalent to \eqref{eq:glm-sparse-condition} up to constants. Moreover, condition (2) is met by assumption. Apply Theorem~\ref{thm:robust_pca} to complete the argument.
\end{proof}

\subsection{Proof of Lemma~\ref{lem:GLM_coherence}}

We now proceed to proving Lemma~\ref{lem:GLM_coherence} and address the three different bounds separately. Suppose that $\mtx{L}_0 = \mtx{S} \mtx{G}^\transp \sim \mathrm{GLM}(\mtx{S},m)$ is sampled from the Gaussian loadings model. Then, $\nu (\mtx{S})$ ensures that $\mtx{S}$ has full column rank, while $m \geq r$ implies that the standard normal matrix $\mtx{G}$ has full column rank with probability one. This ensures that both left- and right incoherence are determined by the individual factors:
\begin{equation} \label{eq:incoherence_factorization}
\mu_{(l)} (\mtx{L}_0) = \mu_{(l)} (\mtx{S}) \quad \textrm{and} \quad \mu_{(r)} (\mtx{L}_0) = \mu_{(r)} (\mtx{G}).
\end{equation}

\subsubsection{Deterministic bound for the left-incoherence}

\begin{lemma} \label{lem:deterministic_incoherence}
A sign matrix $\mtx{S} \in \left\{ \pm 1 \right\}^{n \times r}$ with permeance
$\nu (\mtx{S}) >0$ obeys
\begin{displaymath}
\mu_{(l)} (\mtx{S}) \leq \nu (\mtx{S})^{-1}.
\end{displaymath}
\end{lemma}

\begin{proof}
The function $f: \Sym_n \to \mathbb{R}$ defined by $\mtx{X} \mapsto \max_{1 \leq k \leq n}\langle \mathbf{e}_k, \mtx{X} \mathbf{e}_k \rangle$ is monotone with respect
to the psd order, i.e.\ $\mtx{X} \psdge \mtx{Y}$ implies $f(\mtx{X}) \geq f(\mtx{Y})$.
Next, let $\mtx{P}\in \Sym_n$ denote the projector onto the range of
$\mtx{S}$. 
Then, $\mtx{P} \psdle \lambda^{-1}_{\min\neq 0} \mtx{S} \mtx{S}^\transp$,
where $\lambda_{\min\neq 0}(\mtx{S}\mtx{S}^\transp)$ is the smallest non-zero eigenvalue of $\mtx{S} \mtx{S}^\transp$. Apply an SVD $\mtx{S}=\mtx{U} \mtx{\Sigma} \mtx{V}^\transp$ and use orthogonal invariance of eigenvalues to conclude
\begin{displaymath}
\lambda_{\min\neq 0}(\mtx{S}\mtx{S}^\transp) = \lambda_{\min \neq 0} \left(\mtx{U} \mtx{\Sigma}^2 \mtx{U}^\transp \right)= \sigma_r^2 = \lambda_{\min} \left( \mtx{V} \mtx{\Sigma}^2 \mtx{V}^\transp \right) = \lambda_{\min} \left(\mtx{S}^\transp \mtx{S} \right)= n \nu (\mtx{S}).
\end{displaymath}
Combine this observation with monotonicity of $f: \Sym_n \to \mathbb{R}$:
\begin{displaymath}
\mu_{(l)} (\mtx{S})
= \frac{n}{r} \max_{1 \leq k \leq n} \| \mtx{P} \mathbf{e}_k
\|_{\ell_2}^2
= \frac{n}{r} f (\mtx{P})
\leq \frac{n}{r} f \left(\tfrac{1}{n \nu (\mtx{S})} \mtx{S} \mtx{S}^\transp \right)
=
\frac{1}{r \nu (\mtx{S})} \max_{1 \leq k \leq n} \langle \mathbf{e}_k, \mtx{S}
\mtx{S}^\transp \mathbf{e}_k \rangle.
\end{displaymath}
Finally, note that any sign matrix $\mtx{S} = \begin{bmatrix} \vct{s}_1 & \dots & \vct{s}_r \end{bmatrix} \in \{ \pm 1 \}^{n \times r}$ obeys
$
\langle \mathbf{e}_k, \mtx{S} \mtx{S}^\transp \mathbf{e}_k \rangle
= \sum_{i=1}^r \langle \mathbf{e}_k, \vct{s}_i \rangle^2 = r
$ for all $1 \leq k \leq n$ simultaneously.
\end{proof}

\subsubsection{Probabilistic bound for the right incoherence}

Rel.~\eqref{eq:incoherence_factorization} asserts that the right-incoherence of a Gaussian loadings sample is fully characterized by the standard normal matrix $\mtx{G} \in \mathbb{R}^{m \times r}$. 
Rotation invariance of the standard normal columns extends to the range of $\mtx{G}$. Moreover, with probability one, this range is $r$-dimensional.
Condition on this almost sure event. Then, the projector $\mtx{Q} \in \Sym_m$ on the range of $\mtx{G}$ is a random matrix of the form
\begin{equation} \label{eq:haar_random_subspace}
\mtx{Q}= \mtx{R} \left( \sum_{i=1}^r \mathbf{e}_i \mathbf{e}_i^\transp
\right) \mtx{R}^\transp \quad \textrm{where} \quad \mtx{R} \overset{\textrm{unif.}}{\sim}
\mathsf{O}(m).
\end{equation}
Here $\mtx{R} \overset{\textrm{unif.}}{\sim} \mathsf{O}(m)$ implies that the matrix
$\mtx{R}$ is chosen with respect to the unique invariant measure
on the orthogonal group. This measure is also known as the Haar measure.
For technical reasons, we point out a reformulation of
Eq.~\eqref{eq:haar_random_subspace}. The orthogonal group is a compact Lie group and
therefore unimodular. This implies that the Haar measure is invariant under taking
inverses (transpositions):
\begin{equation} \label{eq:haar_random_subspace_transpose}
\mtx{Q} = \mtx{R}^\transp \left( \sum_{i=1}^r \mathbf{e}_i \mathbf{e}_i^\transp
\right) \mtx{R} \quad \textrm{where} \quad \mtx{R} \overset{\textrm{unif.}}{\sim}
\mathsf{O}(m). 
\end{equation}

\begin{lemma}
Let $\mtx{G} \in \mathbb{R}^{m \times r}$ be a random rank-$r$ matrix with rotation-invariant range in the sense of Eq.~\eqref{eq:haar_random_subspace_transpose}. Then, for any $\alpha >0$
\begin{displaymath}
\mathrm{Pr} \left[ \mu_{(l)} (\mtx{G}) \geq 4 (\alpha+1) \log (n+m) \right] \leq
2(nm)^{-\alpha}/n.
\end{displaymath}
\end{lemma}
\noindent
This result covers Gaussian random matrices, as well as
the random orthogonal model discussed in Ref.~\cite{CLMW11:RobustPCA}.

\begin{proof}
Fix a standard basis vector $\mathbf{e}_{l_0} \in \mathbb{R}^m$ ($1 \leq l_0 \leq m$) and consider the random variable $\langle \mathbf{e}_{l_0},
\mtx{Q} \mathbf{e}_{l_0} \rangle$.
Use rotation invariance \eqref{eq:haar_random_subspace_transpose} to reformulate the distribution of this random variable:
\begin{displaymath}
\langle \mathbf{e}_{l_0}, \mtx{Q} \mathbf{e}_{l_0} \rangle = \langle
\mtx{R} \mathbf{e}_{l_0}, \left( \sum_{i=1}^r \mathbf{e}_i \mathbf{e}_i^\transp \right) \mtx{R}
\mathbf{e}_{l_0} \rangle= \sum_{i=1}^r \langle \mathbf{e}_i, \mtx{R} \mathbf{e}_{l_0} \rangle^2
\quad \textrm{where}\quad \mtx{R} \overset{\textrm{unif}}{\sim} \mathsf{O}(m).
\end{displaymath}
The uniform distribution of $\mtx{R}$ implies that the unit vector $\hat{\vct{v}} = \mtx{R}
\mathbf{e}_{l_0}$ is distributed uniformly over the unit sphere in $\mathbb{R}^m$.
Each component $\langle \mathbf{e}_i, \vct{v} \rangle$ is approximately normal with mean
zero and variance $m^{-1/2}$. This suggests that $\langle \mathbf{e}_{l_0},
\mtx{Q} \mathbf{e}_{l_0}\rangle$ resembles a re-normalized $\chi^2$-distribution with $r$ degrees of
freedom. Lemma~\ref{lem:spherical_tail_bound} below makes this intuition precise and
ensures
\begin{displaymath}
\mathrm{Pr} \left[
\sum_{i=1}^r \langle \vct{e}_i, \hat{\vct{v}} \rangle^2 \geq \tfrac{4 (\alpha+1)r}{m} \log (n+m) \right] \leq 2(nm)^{-(\alpha+1)}.
\end{displaymath}
The final claim follows from a union bound over all $m$ possible choices of the standard basis vector $\vct{e}_{l_0}$.
\end{proof}

\subsubsection{Probabilistic bound for the third incoherence parameter}

\begin{lemma}
Fix $\mtx{S} \in \left\{ \pm 1 \right\}^{n \times r}$ with permeance $\nu (\mtx{S})>0$, $m \geq r$ and sample $\mtx{L}_0\sim \mathrm{GLM}(\mtx{S},m)$ from the Gaussian loadings model. Then, for any $\alpha >0$
\begin{displaymath}
\mathrm{Pr} \left[ \tilde{\mu}(\mtx{L}_0) \geq \tfrac{4(\alpha+1)}{\nu (\mtx{S})} \log (n+m) \right] \leq 2(nm)^{-\alpha}.
\end{displaymath}
\end{lemma}

\begin{proof}
Let $\mtx{L}_0=\mtx{U} \mtx{\Sigma}\mtx{V}^\transp$ be a SVD.
By definition of $\tilde{\mu}(\cdot)$, we must establish
\begin{equation} \label{eq:third_incoherence_bound}
\mathrm{Pr} \left[ \max_{1 \leq k \leq n, 1 \leq l \leq m}\left| \langle
\mathbf{e}_l, \mtx{V} \mtx{U}^\transp \mathbf{e}_k \rangle \right|^2 \geq \tfrac{4(\alpha+1)r}{\nu (\mtx{S})mn} \log
(n+m) \right] \leq 2(nm)^{-\alpha}.
\end{equation}
Fix $\vct{e}_{k_0} \in\mathbb{R}^n$ and $\vct{e}_{l_0} \in \mathbb{R}^m$.
Set $\vct{u} = \mtx{U}^\transp \vct{e}_{k_0} \in \mathbb{R}^r$ and note that its Euclidean length is controlled via left-incoherence and Lemma~\ref{lem:deterministic_incoherence}:
\begin{displaymath}
\| \mtx{U}^\transp \vct{e}_{k_0} \|_{\ell_2}^2 \leq \max_{1 \leq k \leq n} \left\| \mtx{U}^\transp \vct{e}_k \right\|_{\ell_2}^2 = \frac{r}{n} \mu_{(l)} (\mtx{L}_0) = \frac{r}{n} \mu_{(l)}(\mtx{S}) \leq \frac{r}{n \nu (\mtx{S})}.
\end{displaymath}
Let $\hat{\vct{u}}= \mtx{U}^\transp \vct{e}_{k_0}/ \| \mtx{U}^\transp \vct{e}_{k_0} \|_{\ell_2}$ be the unit vector pointing into the same direction as $\vct{u}$.
Then, this bound on the Euclidean length ensures
\begin{displaymath}
\begin{aligned}
\mathrm{Pr} \left[ \left| \langle \vct{e}_{l_0}, \mtx{V} \mtx{U}^\transp \vct{e}_{k_0} \rangle \right|^2 \geq \tfrac{4(\alpha+1)r}{\nu (\mtx{S}) nm} \log (n+m) \right] 
=&  \mathrm{Pr} \left[ \left| \langle \vct{e}_{l_0},\mtx{V} \hat{\vct{u}} \rangle \right|^2 \geq \tfrac{4 (\alpha+1)r}{\nu (\mtx{S}) nm} \| \mtx{U}^\transp \vct{e}_{k_0} \|_{\ell_2}^{-2} \log (n+m) \right] \\
\leq & \mathrm{Pr} \left[ \left| \langle \vct{e}_{l_0}, \mtx{V} \hat{\vct{u}} \rangle \right|^2 \geq \tfrac{4 (\alpha+1)}{m} \log (n+m) \right]
\end{aligned}
\end{displaymath}
The vector $\mtx{V} \hat{\vct{u}} \in \mathbb{R}^m$ also has unit length, because $\mtx{V}$ is orthonormal.
What is more, rotational invariance of the Gaussian matrix $\mtx{G}$ extends
to the matrix of right singular vectors: $\mtx{V}$: $\mtx{V} \sim \mtx{R} \mtx{V}$ for
any $\mtx{R} \in \mathsf{O}(m)$.
This implies that $\mtx{V} \hat{\vct{u}}$ is distributed uniformly on the $m$-dimensional unit
sphere. Lemma~\ref{lem:spherical_tail_bound} then implies
\begin{displaymath}
\mathrm{Pr} \left[ | \langle \mathbf{e}_{l_0}, \mtx{V} \hat{\vct{u}} \rangle |^2 \geq 
\tfrac{4(\alpha+1)}{m}\log (n+m)\right] \leq 2(nm)^{-(\alpha+1)}.
\end{displaymath}
Finally, deduce the claim by applying a union bound over all 
$nm$ choices of standard basis vectors $\mathbf{e}_{k_0} \in \mathbb{R}^n$ and $\mathbf{e}_{l_0} \in\mathbb{R}^m$.
\end{proof}

\subsection{Exponential tail bound for vectors chosen uniformly from the unit sphere}

\begin{lemma} \label{lem:spherical_tail_bound}
Let $\left\{\vct{u}_1,\ldots,\vct{u}_r \right\} \subset \mathbb{R}^m$ be a set of unit vectors and
choose $\hat{\vct{v}}$ uniformly from the unit sphere in $m$ dimensions.
Then, for any $t >0$
\begin{displaymath}
\mathrm{Pr} \left[ \sum_{i=1}^r \langle \vct{u}_i, \hat{\vct{v}} \rangle^2 \geq\tfrac{4rt}{m}
\right] \leq 2\mathrm{e}^{-t}.
\end{displaymath}
\end{lemma}
\noindent
This is a typical tail bound. It follows from establishing sub-exponential moment
growth and subsequently applying the exponential Markov inequality.
Stronger concentration inequalities follow from concentration of measure. However, these tighter bounds are somewhat unwieldy by comparison.

\begin{proof}
Define the random variable $S = \tfrac{m}{r} \sum_{i=1}^r \ip{\vct{u}_i}{\hat{\vct{v}}}^2$ and interpret it
as the squared Euclidean norm of a random vector $\tilde{\vct{v}}=\sqrt{\tfrac{m}{r}} \begin{bmatrix} \ip{\vct{u}_1}{\hat{\vct{v}}}& \ldots & \ip{\vct{u}_1}{\hat{\vct{v}}} \end{bmatrix}^\transp \in \mathbb{R}^r$.
The fundamental $\ell_p$-norm relation relation $\| \tilde{\vct{v}} \|_{\ell_2} \leq r^{1/2-1/(2p)} \| \tilde{\vct{v}}
\|_{\ell_{2p}}$ in $\mathbb{R}^r$ then implies
\begin{displaymath}
\mathbb{E} \left[ S^p \right] = \mathbb{E} \left[ \| \tilde{\vct{v}} \|_{\ell_2}^{2p}
\right]
\leq r^{p-1} \mathbb{E} \left[ \| \tilde{\vct{v}} \|_{\ell_{2p}}^{2p} \right]
= \frac{m^p}{r} \sum_{i=1}^r \mathbb{E} \left[ \langle \vct{u}_i, \vct{v} \rangle^{2p}
\right] \quad \textrm{for any} \quad p \in \mathbb{N}.
\end{displaymath}
Next, let $\vct{g} \in \mathbb{R}^m$ be a standard normal
vector. This random vector may be decomposed into a direction
$\hat{\vct{v}} = \vct{g} / \| \vct{v} \|_{\ell_2}$ and a radius $\rho = \| \vct{g} \|_{\ell_2}$.
The direction is distributed uniformly on the $m$-dimensional unit sphere and $\rho^2$
follows a $\chi^2$-distribution with $m$ degrees of freedom. Importantly, $\hat{\vct{v}}$ and
$\rho$ are stochastically independent. Combine this with rotation invariance and normalization ($\| \vct{u}_i \|_{\ell_2}=1$) to conclude
\begin{align*}
\mathbb{E} \left[ \langle \vct{u}_i, \hat{\vct{v}} \rangle^{2p} \right]
= \frac{ \mathbb{E} \left[ \langle \vct{u}_i, \vct{g} \rangle^{2p}\right]}{\mathbb{E}
\left[ \| \vct{g} \|_{\ell_2}^{2p} \right]}
= \frac{ \| \vct{u}_i \|_{\ell_2}^{2p} \mathbb{E} \left[ \langle \mathbf{e}_1, \vct{g}
\rangle^{2p}\right]}{\mathbb{E} \left[ (\chi^2_m)^p\right]}
= \frac{ (2p-1)!!}{m (m+2) \cdots (m+2(p-1))}\quad \textrm{for all} \quad 1 \leq i \leq r.
\end{align*}
The last equation follows from well-known expressions for
the moments of standard normal and $\chi^2_m$-distributed random variables.
The relations $(2p-1)!! \leq (2p)!!=2^p p!$ and $m(m+2) \cdots (m+2(p-1)) \geq m^p$ then
imply
\begin{displaymath}
\mathbb{E} \left[ S^p \right] \leq \frac{m^p}{r} \sum_{i=1}^r \frac{2^p p!}{m^p} = 2^p
p!\quad \textrm{for all} \quad p \in \mathbb{N}.
\end{displaymath}
This moment growth suggests sub-exponential tail behavior. The exponential Markov inequality makes this intuition precise:
\begin{displaymath}
\begin{aligned}
\mathrm{Pr} \left[ \sum_{i=1}^r \langle \vct{u}_i, \hat{\vct{v}} \rangle^2 \geq  \tfrac{4rt}{m}
\right]
=& \mathrm{Pr} \left[ \tfrac{S}{4} \geq t \right] = \mathrm{Pr} \left[ \mathrm{e}^{S/4} \geq
\mathrm{e}^t \right] \leq \mathrm{e}^{-t} \mathbb{E} \left[ \exp \left( \tfrac{S}{4} \right)
\right] \\
=& \mathrm{e}^{-t} \sum_{p=0}^\infty \frac{ \mathbb{E} \left[ S^p \right]}{4^p p!} \leq
\mathrm{e}^{-t} \sum_{p=0}^\infty 2^{-p} = 2 \mathrm{e}^{-t}.
\end{aligned}
\end{displaymath}
\end{proof}

\section{Gaussian loadings model: Removing gross errors in matrix columns}
\label{app:glm-outliers}

\subsection{Background}

We consider a noise model in which a data matrix is corrupted by a potentially large number of outliers \cite{HR09:Robust-Statistics, LMTZ15:Robust-Computation}. Consider a data matrix that consists of $m$ low-dimensional inliers and $m'$
high-dimensional outliers. More precisely, set $M=m+m'$ and consider a matrix of the form
\begin{equation}\label{eq:in_and_out_model}
\mtx{B}
= \begin{bmatrix} \mtx{L}_0  &  
\mtx{\Omega}_0 \end{bmatrix} \mtx{\Pi} \in \mathbb{R}^{n \times M}.
\end{equation}
Here, $\mtx{L}_0 \in \mathbb{R}^{n \times m}$ is a rank-$r$ matrix of clean data, $\mtx{\Omega}_0 \in \mathbb{R}^{n \times m'}$ subsumes arbitrary noise and $\mtx{\Pi} \in \mathbb{R}^{M \times M}$ is an unknown permutation.
We do not know $m$ (number of inliers) and $m'$ (number of column outliers) in advance, but we require knowledge of $r = \mathrm{rank}(\mtx{L}_0)$.
This In\&Out model has a key feature. The inliers (columns of $\mtx{L}_0$) are confined to the $r$-dimensional range $\mathsf{L}_0 \subset \mathbb{R}^n$ of $\mtx{L}_0$, while the outliers can be arbitrary.
Hence, the projector $\mtx{P}_0 \in \Sym_n$ onto $\mathsf{L}_0$ discriminates inliers from outliers.
The \emph{reaper} algorithm \cite{LMTZ15:Robust-Computation} is designed to recover this projector from the corrupted data matrix:
\begin{equation} \label{eq:reaper-appendix}
\begin{aligned}
&\underset{\mtx{P} \in \Sym_n}{\minimize} &&
\sum_{i=1}^{M} \norm{ (\Id - \mtx{P}) \mtx{B} \mathbf{e}_i }_{\ell_2}
&\subjto &&
\text{$\trace(\mtx{P}) = r$ and $\mtx{0} \psdle \mtx{P} \psdle \Id$.}
\end{aligned}
\end{equation}
Sufficient conditions for exact recovery have been established by Lerman, McCoy, Tropp and Zhang
\cite{LMTZ15:Robust-Computation}.
These criteria are deterministic and depend on data-dependent parameters that play a
role analogous to the incoherence statistics discussed in Appendix~\ref{app:rank-sparsity}.
Define the \emph{permeance statistics} for
the clean data matrix  $\mtx{L}_0 = \begin{bmatrix} \vct{l}_1 & \ldots &\vct{l}_m \end{bmatrix}$ with range $\mathsf{L}_0$:
\begin{equation} \label{eq:permeance-statistics}
P \left( \mathsf{L}_0, \mtx{L}_0\right)
= \inf_{\vct{u} \in \mathsf{L}_0, \|\vct{u} \|_{\ell_2}=1} \sum_{i=1}^{m} \left|
\langle \vct{u}, \vct{l}_i\rangle \right|.
\end{equation}

\begin{remark}[Relation between permeance statistics and permeance parameter] \label{rem:permeance} Permeance parameter \eqref{eq:permeance-appendix} and permeance statistics \eqref{eq:permeance-statistics} are closely related. The permeance statistics minimizes a $\ell_1$-penalty, while the permeance parameter optimizes the least-squares loss of the same objective:
\begin{displaymath}
n \nu (\mtx{S}) = \lambda_{\min}(\mtx{S}^\transp \mtx{S})
= \inf_{\vct{u} \in \mathrm{ran}(\mtx{S}),\| \vct{u} \|_{\ell_2}=1} \left\| \mtx{S} \vct{u} \right\|_{\ell_2}^2 
= \inf_{\vct{u} \in \mathrm{ran}(\mtx{S}),\| \vct{u} \|_{\ell_2}=1} \sum_{i=1}^r \langle \vct{u}, \vct{s}_i \rangle^2.
\end{displaymath}
\end{remark}
\noindent
Define the \emph{spherical linear structure statistics} for the matrix of outliers $\mtx{\Omega}_0 = \begin{bmatrix} \vct{\omega}_1 & \ldots & \vct{\omega}_{m'} \end{bmatrix}$ and the ortho-complement $\mathsf{L}_0^\perp$ of $\mathsf{L}_0$:
\begin{displaymath}
\hat{S}\left( \mathsf{L}_0^\perp, \mtx{\Omega}_0 \right)^2
= \sup_{\vct{u} \in \mathsf{L}_0^\perp, \| \vct{u} \|_{\ell_2}=1} \sum_{i=1}^{m'} \left| \ip{ \vct{u}}{\widehat{\mtx{P}_{\mathsf{L}_0^\perp} \vct{\omega}_i}} \right|^2.
\end{displaymath}
Here, $\mtx{P}_{\mathsf{L}_0^\perp} \in \Sym_n$ denotes the orthogonal projector onto $\mathsf{L}_0^\perp$ and $\hat{\vct{x}}=\vct{x}/\| \vct{x} \|_{\ell_2}$ denotes the unit vector pointing into the same direction as $\vct{x}$ (with the convention that $\hat{\vct{0}}=\vct{0}$).
Finally, let $\| \cdot \|_{S_\infty}$ denote the spectral (or operator) norm. The main result \cite[Thm.~3.1]{LMTZ15:Robust-Computation} establishes a deterministic recovery guarantee for the reaper algorithm \eqref{eq:reaper-appendix} based on these parameters.

\begin{theorem}[Lerman, McCoy, Tropp \& Zhang] \label{thm:reaper}
\label{thm:reaper_deterministic}
Consider a data matrix $\mtx{B}=\begin{bmatrix} \mtx{L}_0 & \mtx{\Omega}_0 \end{bmatrix} \mtx{\Pi} \in \mathbb{R}^{n \times M}$ and set $\mathsf{L}_0 = \mathrm{ran}(\mtx{L}_0) \subset \mathbb{R}^n$.
Suppose that 
\begin{displaymath}
P \left( \mathsf{L}_0, \mtx{L}_0\right)
> \sqrt{2r} \hat{S}\left( \mathsf{L}_0^\perp, \mtx{\Omega}_0 \right) \left\| \mtx{\Omega}_0 \right\|_{S_\infty}.
\end{displaymath}
Then, the solution to the reaper problem \eqref{eq:reaper-appendix} is the orthogonal projection $\mtx{P}_0$ onto $\mathsf{L}_0$.
\end{theorem}

\subsection{Summary statistics parameters for the Gaussian loadings model and denoising}

We consider a clean data matrix sampled from the Gaussian loadings model (Model~\ref{mod:glm}):
\begin{displaymath}
\mtx{L}_0 = \tfrac{1}{\sqrt{r}} \mtx{S} \mtx{G}^\transp \quad \textrm{where} \quad \mtx{S} \in \left\{ \pm 1 \right\}^{n \times r} \quad \textrm{and} \quad \mtx{G} \in \mathbb{R}^{m \times r} \quad \textrm{has standard normal entries}.
\end{displaymath}
The relation between permeance statistics and permeance parameter displayed in Remark~\ref{rem:permeance} suggests that the former is controlled by the square root of the latter. The following technical results makes this intuition precise.

\begin{lemma} \label{lem:permeance-bound}
Let $\mathsf{L}_0 \in \mathbb{R}^n$ denote the range of a sign matrix $\mtx{S} \in \left\{\pm 1 \right\}^{n \times r}$ with permeance $\nu (\mtx{S})$.
A sample $\mtx{L}_0 \sim \mathrm{GLM}(\mtx{S},m)$ from the associated Gaussian loadings model obeys
\begin{displaymath}
\mathrm{Pr} \left[ P(\mathsf{L}_0,\mtx{L}_0) \geq \tfrac{1}{6} \sqrt{\tfrac{nm}{r}} \left( \sqrt{m \nu (\mtx{S})} - 12 \sqrt{r} - 2 \sqrt{\nu (\mtx{S})} t \right) \right]
\geq  1-\mathrm{e}^{-t^2/2} \quad \textrm{for any} \quad t >0.
\end{displaymath}
\end{lemma}
\noindent
This claim follows from a versatile proof technique known as Mendelson's small ball method \cite{Men15:Small-Ball,KM15:Small-Ball,Tro15:Bowling}. We present a detailed proof in the next sub-section. 

Next, we consider the impact of white noise corruptions. Let $\mtx{\Omega}_0 \in \mathbb{R}^{n \times m'}$ be a random matrix with standard normal entries. 
The spectral properties of such random matrices are very well understood:
\begin{equation}\label{eq:Gaussian-Norm}
\mathrm{Pr} \left[ \left\| \mtx{\Omega}_0 \right\|_{S_\infty} \leq \sqrt{n} + \sqrt{m'} +t \right] \geq 1 - \mathrm{e}^{-t^2} \quad \textrm{for any} \quad t >0,
\end{equation}
see e.g.\ \cite[Theorem~2.13]{DKS01:RandomMatrices}.
Similar results remain valid for random matrices whose columns are sampled uniformly from the $m'$-dimensional unit sphere instead. We borrow the following extension from Lerman et al.

\begin{lemma}[Lem.~8.4 in \cite{LMTZ15:Robust-Computation}] \label{lem:spherical-bound}
Fix a $r$-dimensional subspace $\mathsf{L}_0 \in \mathbb{R}^n$. Then, a standard Gaussian matrix $\mtx{\Omega}_0 \in \mathbb{R}^{n \times m'}$ obeys
\begin{displaymath}
\mathrm{Pr} \left[ \hat{S} \left( \mathsf{L}_0^\perp, \mtx{\Omega}_0 \right) \leq \frac{\sqrt{m'}+\sqrt{n-r}+t}{\sqrt{n-r-0.5}} \right] \geq 1 - 1.5 \mathrm{e}^{-t^2/2}
\end{displaymath}
\end{lemma}
\noindent
We may insert these probabilistic bounds into Theorem~\ref{thm:reaper}. 
Applying the statement then ensures that the reaper algorithm \eqref{eq:reaper-appendix} is capable of perfectly removing random white noise corruptions.

\begin{theorem}[Theorem~\ref{thm:glm-outliers}, restatement]
Suppose we observe the matrix
$\mtx{B} = \begin{bmatrix} \mtx{L}_0 & \mtx{\Omega}_0 \end{bmatrix} \mtx{\Pi}
\in \R^{n \times M}$
where $M = m + m'$ and
\begin{enumerate}
\item	The sign matrix $\mtx{S} \in \{\pm 1\}^{n \times r}$ is Schur independent, with permeance $\nu(\mtx{S})$;
\item	The data matrix $\mtx{L}_0 \sim \mathrm{GLM}(\mtx{S}, m)$ is drawn from the Gaussian loadings model; 
\item	The noise matrix $\mtx{\Omega}_0 \in \R^{n \times m'}$ is standard normal; and
\item	The permutation matrix $\mtx{\Pi} \in \R^{M \times M}$ is arbitrary.
\end{enumerate}
Assume that the parameters satisfy
\begin{equation} \label{eq:reaper-cond-appendix}
\max\left\{ 1, \frac{m'}{n} \right\}
\lesssim \sqrt{ \nu(\mtx{S}) } \cdot \frac{m}{r}.
\end{equation}
Then, with high probability over the randomness in the model,
the orthogonal projector onto $\mathsf{L}_0=\range(\mtx{L}_0)$ is the unique
solution to the optimization problem~\eqref{eq:reaper-appendix}.
\end{theorem}

\begin{proof}
Choose $t \propto \min \left\{m,m'\right\}$ and invoke Lemmas \ref{lem:permeance-bound}--\ref{lem:spherical-bound} and Rel.~\eqref{eq:Gaussian-Norm} to ensure
\begin{displaymath}
P\left(\mathsf{L}_0,\mtx{L}_0 \right) \gtrsim m \sqrt{\tfrac{n \nu (\mtx{S})}{r}} \quad \textrm{and} \quad
\hat{S}\left(\mathsf{L}_0^\perp, \mtx{\Omega}_0 \right) \| \mtx{\Omega}_0 \|_{S_\infty} \lesssim \max \left\{ \sqrt{n}, \tfrac{m'}{\sqrt{n}}\right\}
\end{displaymath}
with high probability each. Condition on these bounds to be valid.
Then, Rel.~\eqref{eq:reaper-cond-appendix} implies the deterministic recovery condition from Theorem~\ref{thm:reaper}. Applying this statement ensures correct recovery of $\mtx{P}_0$ via the reaper algorithm. This argument is valid, irrespective of the specific choice of perturbation $\mtx{\Pi}$.
\end{proof}

\subsection{Proof of Lemma~\ref{lem:permeance-bound}}

We use Mendelson's small ball method to derive a strong probabilistic lower bound on the permeance statistics associated with the Gaussian loadings model.
The following variant of this versatile technique can be extracted from Tropp's proof \cite{Tro15:Bowling}.

\begin{theorem}[Mendelson's small ball method]
\label{thm:mendelson}
Fix a set $\mathsf{E} \subset \mathbb{R}^n$ and let $\vct{\phi}_1,\ldots,\vct{\phi}_m$
be independent copies of a random vector $\vct{\phi} \in\mathbb{R}^n$. Fix $\xi >0$ and
define
\begin{displaymath}
\begin{aligned}
Q_{\xi} \left( \mathsf{E};\vct{\phi} \right) =& \inf_{\vct{u} \in \mathsf{E}}
\mathrm{Pr} \left[ \left| \langle \vct{u},\vct{\phi} \rangle \right| \geq \xi \right] &
\textrm{(marginal tail function)},  \\
W_m \left( \mathsf{E},\vct{\phi} \right) =& \mathbb{E} \sup_{\vct{u} \in \mathsf{E}} \langle
\vct{u}, \vct{h} \rangle \quad \textrm{with} \quad  \vct{h} = \frac{1}{\sqrt{m}} \sum_{i=1}^m \varepsilon_i
\vct{\phi}_i & \textrm{(empirical width)},
\end{aligned}
\end{displaymath}
where $\varepsilon_1,\ldots,\varepsilon_m \in \left\{ \pm 1 \right\}$ are independent
Rademacher\footnote{A Rademacher random variable takes the values $+1$ and $-1$ with
equal probability.} random variables.
Then, for any $t >0$,
\begin{displaymath}
\mathrm{Pr} \left[ \tfrac{1}{\sqrt{m}} \inf_{\vct{u} \in \mathsf{E}} \sum_{i=1}^m \left|
\langle \vct{u}, \vct{\phi}_i \rangle \right| \geq \xi \sqrt{m} Q_{2 \xi} (\mathsf{E};
\vct{\phi}) - 2 W_m (\mathsf{E},\phi) - \xi t \right] \geq 1- \mathrm{e}^{-t^2/2}.
\end{displaymath}
\end{theorem}
\noindent
The original formulation \cite{Men15:Small-Ball, KM15:Small-Ball} lower-bounds the minimal $\ell_2$-loss
and readily follows from Lyapunov's inequality.

Lemma~\ref{lem:permeance-bound} follows from adapting Theorem~\ref{thm:mendelson} to the Gaussian loadings model.
Two technical lemmas are instrumental for controlling the empirical width and the marginal tail function.

\begin{lemma} \label{lem:empirical-width}
Fix a sign matrix $\mtx{S} \in \left\{ \pm 1 \right\}^{n \times r}$, $m$ independent standard Gaussian vectors $\vct{g}_1,\ldots,\vct{g}_m \in \mathbb{R}^r$ and $m$ independent Rademacher random variables $\epsilon_1,\ldots,\epsilon_m$. 
Then,
\begin{displaymath}
\mathbb{E} \sup_{\hat{\vct{u}} \in \mathrm{ran}(\mtx{S}),\|\hat{\vct{u}} \|_{\ell_2}=1} \ip{\hat{\vct{u}}}{\tfrac{1}{\sqrt{m}} \sum_{i=1}^m \epsilon_i \mtx{S} \vct{g}_i} \leq \sqrt{n}.
\end{displaymath}
\end{lemma}

\begin{proof}
Note that the vector $\vct{h} = \tfrac{1}{\sqrt{m}}\sum_{i=1}^m \epsilon_i \mtx{S} \vct{g}_i$ is necessarily contained in the range of the sign matrix $\mtx{S}$. Combine this insight with Jensen's inequality and independence of the Rademacher random variables ($\mathbb{E} \epsilon_i \epsilon_j = \delta_{i,j}$):
\begin{displaymath}
\mathbb{E} \sup_{\vct{u} \in \mathrm{ran}(\mtx{S}),\|\vct{u} \|_{\ell_2}=1} \ip{\vct{u}}{\vct{h}}= \mathbb{E} \| \vct{h} \|_{\ell_2} \leq \left( \mathbb{E} \| \vct{h} \|_{\ell_2}^2 \right)^{1/2} = \left(\frac{1}{m} \sum_{i=1}^m \mathbb{E} \langle \mtx{S} \vct{g}_i, \mtx{S} \vct{g}_i \rangle \right)^{1/2} = \left( \mathbb{E} \langle \mtx{S} \vct{g},\mtx{S} \vct{g} \rangle \right)^{1/2}.
\end{displaymath}
Finally, we apply isotropy of standard normal vectors ($\mathbb{E} \vct{g}\vct{g}^\transp=\Id$) to conclude
\begin{displaymath}
\mathbb{E} \langle \mtx{S} \vct{g}, \mtx{S} \vct{g} \rangle = \tfrac{1}{r} \mathbb{E} \ip{\vct{g}}{\mtx{S}^\transp \mtx{S} \vct{g}}=\tfrac{1}{r} \mathrm{tr} \left( \mtx{S}^\transp \mtx{S} \right)=n.
\end{displaymath}
\end{proof}

\begin{lemma}\label{lem:marginal-tail}
Fix a sign matrix $\mtx{S} \in \left\{ \pm 1 \right\}^{n \times r}$ with permeance $\nu (\mtx{S})$ and sample a standard normal vector $\vct{g} \in \mathbb{R}^r$. Then,
\begin{displaymath}
\mathrm{Pr} \left[ \left| \ip{\hat{\vct{u}}}{\mtx{S} \vct{g}} \right| \geq \tfrac{2}{3}\sqrt{\nu (\mtx{S}) n} \right] \geq \tfrac{1}{2} \quad \textrm{for all unit vectors} \quad \hat{\vct{u}} \in \mathrm{ran}(\mtx{S}).
\end{displaymath}
\end{lemma}

\begin{proof}
Use the reformulation of $\nu (\mtx{S})$ provided in Remark~\ref{rem:permeance} to infer
$
\tfrac{2}{3} \sqrt{\nu (\mtx{S})n} 
\leq \tfrac{2}{3} \| \mtx{S}^\transp \hat{\vct{u}} \|_{\ell_2}
$ for all unit vectors $\hat{\vct{u}} \in \mathrm{ran}(\mtx{S})$.
Combine this with rotation invariance to deduce 
\begin{displaymath}
\mathrm{Pr} \left[ \left| \ip{\hat{\vct{u}}}{\mtx{S} \vct{g}} \right| \geq 2\sqrt{\tfrac{\nu (\mtx{S}) n}{9r}} \right]
\geq \mathrm{Pr} \left[ \left| \langle \mtx{S}^\transp \hat{\vct{u}}, \vct{g} \rangle \right| \geq \tfrac{2}{3} \| \mtx{S}^\transp \hat{\vct{u}} \|_{\ell_2} \right]
= \mathrm{Pr} \left[ \left| \ip{\mathbf{e}_1}{\vct{g}}\right| \geq \tfrac{2}{3} \right]\quad \textrm{for all} \quad \hat{\vct{u}} \in \mathrm{ran}(\mtx{S}).
\end{displaymath}
Well-known results state that the median of the half-normal random variable $| \ip{\mathbf{e}_1}{\vct{g}}|$ is $\sqrt{2} \mathrm{erf}^{-1} (1/2) > \tfrac{2}{3}$. Apply the defining property of the median to deduce the claim.
\end{proof}

\begin{proof}[Proof of Lemma~\ref{lem:permeance-bound}]
Fix a sign matrix $\mtx{S} \in \left\{ \pm 1 \right\}^{n \times r}$, a natural number $m \geq r$ and suppose that 
$
\mtx{L}_0 =\begin{bmatrix} \vct{l}_1&\ldots & \vct{l}_{m} \end{bmatrix} \sim \mathrm{GLM}(\mtx{S},m)
$
 is sampled from the Gaussian loadings model. By assumption, each column $\vct{l}_i \in \mathbb{R}^n$ ($1 \leq i \leq m$) is an independent copy of the random vector
\begin{displaymath}
\vct{l} = \tfrac{1}{\sqrt{r}} \mtx{S} \vct{g} \quad \textrm{where} \quad \vct{g} \in\mathbb{R}^r \quad \textrm{is standard normal}. 
\end{displaymath}
Set $\mathsf{L}_0 = \mathrm{ran}(\mtx{L}_0) \subset \mathbb{R}^n$ and denote its intersection with the unit sphere by $\mathsf{E}=\left\{ \hat{\vct{u}} \in \mathsf{L}_0:\; \| \hat{\vct{u}} \|_{\ell_2}=1 \right\}$. This allows us to rewrite the permeance statistics \eqref{eq:permeance-statistics} as
\begin{displaymath}
P(\mathsf{L}_0, \mtx{L}_0) = \inf_{\hat{\vct{u}} \in \mathsf{E}} \sum_{i=1}^m \left| \langle \hat{\vct{u}},\vct{l}_i \rangle \right|.
\end{displaymath}
Theorem~\ref{thm:mendelson} lower-bounds expressions of precisely this form.
Note that $\mathsf{L}_0 \subseteq \mathrm{ran}(\mtx{S})$, regardless of the value of $m$.
 Lemma~\ref{lem:empirical-width} then asserts
\begin{displaymath}
W_m \left(\mathsf{E}, \vct{l} \right)
\leq\mathbb{E} \sup_{\hat{\vct{u}} \in \mathrm{ran}(\mtx{S}),\|\hat{\vct{u}} \|_{\ell_2}=1} \ip{\hat{\vct{u}}}{\tfrac{1}{\sqrt{m}} \sum_{i=1}^m \epsilon_i \mtx{S} \vct{g}_i} \leq \sqrt{n},
\end{displaymath}
while Lemma~\ref{lem:marginal-tail} ensures that for $\xi_0 = \sqrt{\tfrac{\nu (\mtx{S})n}{9r}}$, the marginal tail function obeys
\begin{displaymath}
Q_{2 \xi_0} (\mathsf{E},\vct{l}) = \inf_{\hat{\vct{u}} \in \mathsf{E}} \mathrm{Pr} \left[ \left|\ip{\hat{\vct{u}}}{\vct{l}}\right| \geq \tfrac{2}{3 \sqrt{r}} \sqrt{\nu (\mtx{S}) n} \right]
\geq \inf_{\hat{\vct{u}} \in \mathrm{ran}(\mtx{S}),\| \hat{\vct{u}}\|_{\ell_2}=1} \mathrm{Pr} \left[ | \ip{\hat{\vct{u}}}{\mtx{S} \vct{g}}| \geq \tfrac{2}{3} \sqrt{\nu (\mtx{S}) n}\right] \geq \tfrac{1}{2}.
\end{displaymath}
Insert these bounds into the assertion of Theorem~\ref{thm:mendelson} to complete the proof.
\end{proof}

\section*{Acknowledgments}

This research was partially funded by ONR awards N00014-11-1002, N00014-17-12146, and
N00014-18-12363.  Additional support was provided by the Gordon \& Betty Moore Foundation.

\bibliographystyle{myalpha}
\bibliography{binfactor}

\end{document}

%% file: KT19_macro.tex
\usepackage[usenames,dvipsnames]{xcolor}
\usepackage{fancyhdr}
\usepackage{amsmath,amsfonts,amsbsy,amsgen,amscd,mathrsfs,amssymb,amsthm}
\usepackage{subfig}
\usepackage{url}
\usepackage{mathtools}

\mathtoolsset{showonlyrefs}

\usepackage[font=small,margin=0.25in,labelfont={sc},labelsep={colon}]{caption}

\usepackage{tikz}
\usetikzlibrary{positioning}
\usepackage{pgfplots}
\usepackage{microtype}
\usepackage{enumitem}

\definecolor{dark-gray}{gray}{0.3}
\definecolor{dkgray}{rgb}{.4,.4,.4}
\definecolor{dkblue}{rgb}{0,0,.5}
\definecolor{medblue}{rgb}{0,0,.75}
\definecolor{rust}{rgb}{0.5,0.1,0.1}

\usepackage[colorlinks=true]{hyperref}

\hypersetup{urlcolor=rust}
\hypersetup{citecolor=dkblue}
\hypersetup{linkcolor=dkblue}

\usepackage{setspace}

\usepackage{graphicx}
\usepackage{booktabs,longtable,tabu} 
\usepackage{multirow} 

\usepackage{float}

\usepackage{fourier}
\usepackage{bm}

\graphicspath{{art/}}

\newtheorem{bigthm}{Theorem}

\newtheorem{theorem}{Theorem}[section]
\newtheorem{lemma}[theorem]{Lemma}

\newtheorem{proposition}[theorem]{Proposition}
\newtheorem{fact}[theorem]{Fact}

\newtheorem*{lemma*}{Lemma}

\theoremstyle{definition}

\newtheorem{definition}[theorem]{Definition}
\newtheorem{example}[theorem]{Example}
\newtheorem{remark}[theorem]{Remark}

\newtheorem{model}[theorem]{Model}

\theoremstyle{remark}

\newtheorem{problem}[theorem]{Problem}

\usepackage[noend]{algpseudocode}
\usepackage{algorithm,algorithmicx}

\algrenewcommand\alglinenumber[1]{\sf\scriptsize\color{blue}{#1}}
\algrenewcommand\algorithmicrequire{\textbf{Input:}}
\algrenewcommand\algorithmicensure{\textbf{Output:}}

\numberwithin{equation}{section} 
\numberwithin{figure}{section}
\numberwithin{table}{section}

\floatstyle{plaintop}
\newfloat{recipe}{thp}{lor}
\floatname{recipe}{Recipe}
\numberwithin{recipe}{section}

\providecommand{\mathbold}[1]{\bm{#1}}  
                                
\renewcommand{\phi}{\varphi}

\newcommand{\Id}{\mathbf{I}}

\providecommand{\mathbbm}{\mathbb} 

\newcommand{\R}{\mathbbm{R}}

\newcommand{\Sym}{\mathbb{H}}

\newcommand{\Expect}{\operatorname{\mathbb{E}}}

\newcommand{\vct}[1]{\mathbold{#1}}
\newcommand{\mtx}[1]{\mathbold{#1}}

\newcommand{\transp}{\mathsf{t}}

\newcommand{\lspan}{\operatorname{span}}

\newcommand{\range}{\operatorname{range}}

\newcommand{\rank}{\operatorname{rank}}

\newcommand{\diag}{\operatorname{diag}}
\newcommand{\trace}{\operatorname{trace}}

\newcommand{\psdle}{\preccurlyeq}
\newcommand{\psdge}{\succcurlyeq}

\newcommand{\psdgt}{\succ}

\newcommand{\ip}[2]{\langle {#1},\ {#2} \rangle}

\newcommand{\norm}[1]{\left\Vert {#1} \right\Vert}

\newcommand{\minimize}{\mathrm{minimize}}
\newcommand{\maximize}{\mathrm{maximize}}
\newcommand{\subjto}{\mathrm{subject\ to}}

